%
%
%
\def\unredoffs{} 

%
%
%
%
\newbox\leftpage \newdimen\fullhsize \newdimen\hstitle \newdimen\hsbody
\tolerance=1000\hfuzz=2pt
\catcode`\@=11 
\magnification=1200\unredoffs\baselineskip=16pt plus 2pt minus 1pt
\hsbody=\hsize \hstitle=\hsize 
%
%
%
%
%
\newcount\yearltd\yearltd=\year\advance\yearltd by -1900

\def\Title#1#2{\nopagenumbers\abstractfont\hsize=\hstitle\rightline{#1}%
\vskip 1in\centerline{\titlefont #2}\abstractfont\vskip .5in\pageno=0}
\def\Date#1{\vfill\leftline{#1}\tenpoint\supereject\global\hsize=\hsbody%
\footline={\hss\tenrm\folio\hss}}
%

\def\draftmode{\message{ DRAFTMODE }\def\draftdate{{\rm preliminary draft:
\number\month/\number\day/\number\yearltd\ \ \hourmin}}%
\headline={\hfil\draftdate}\writelabels\baselineskip=20pt plus 2pt minus 2pt
 {\count255=\time\divide\count255 by 60 \xdef\hourmin{\number\count255}
  \multiply\count255 by-60\advance\count255 by\time
  \xdef\hourmin{\hourmin:\ifnum\count255<10 0\fi\the\count255}}}
\def\nolabels{\def\wrlabeL##1{}\def\eqlabeL##1{}\def\reflabeL##1{}}
\def\writelabels{\def\wrlabeL##1{\leavevmode\vadjust{\rlap{\smash%
{\line{{\escapechar=` \hfill\rlap{\sevenrm\hskip.03in\string##1}}}}}}}%
\def\eqlabeL##1{{\escapechar-1\rlap{\sevenrm\hskip.05in\string##1}}}%
\def\reflabeL##1{\noexpand\llap{\noexpand\sevenrm\string\string\string##1}}}
\nolabels
%
\global\newcount\secno \global\secno=0
\global\newcount\meqno \global\meqno=1
\def\newsec#1{\global\advance\secno by1\message{(\the\secno. #1)}
\global\subsecno=0\eqnres@t\noindent{\bf\the\secno. #1}
\writetoca{{\secsym} {#1}}\par\nobreak\medskip\nobreak}
\def\eqnres@t{\xdef\secsym{\the\secno.}\global\meqno=1\bigbreak\bigskip}
\def\sequentialequations{\def\eqnres@t{\bigbreak}}\xdef\secsym{}
\global\newcount\subsecno \global\subsecno=0
\def\subsec#1{\global\advance\subsecno by1\message{(\secsym\the\subsecno. #1)}
\ifnum\lastpenalty>9000\else\bigbreak\fi
\noindent{\it\secsym\the\subsecno. #1}\writetoca{\string\quad 
{\secsym\the\subsecno.} {#1}}\par\nobreak\medskip\nobreak}
\def\appendix#1#2{\global\meqno=1\global\subsecno=0\xdef\secsym{\hbox{#1.}}
\bigbreak\bigskip\noindent{\bf Appendix #1. #2}\message{(#1. #2)}
\writetoca{Appendix {#1.} {#2}}\par\nobreak\medskip\nobreak}
%
%
\def\eqnn#1{\xdef #1{(\secsym\the\meqno)}\writedef{#1\leftbracket#1}%
\global\advance\meqno by1\wrlabeL#1}
\def\eqna#1{\xdef #1##1{\hbox{$(\secsym\the\meqno##1)$}}
\writedef{#1\numbersign1\leftbracket#1{\numbersign1}}%
\global\advance\meqno by1\wrlabeL{#1$\{\}$}}
\def\eqn#1#2{\xdef #1{(\secsym\the\meqno)}\writedef{#1\leftbracket#1}%
\global\advance\meqno by1$$#2\eqno#1\eqlabeL#1$$}
%
\newskip\footskip\footskip14pt plus 1pt minus 1pt 
\def\footnotefont{\ninepoint}\def\f@t#1{\footnotefont #1\@foot}
\def\f@@t{\baselineskip\footskip\bgroup\footnotefont\aftergroup\@foot\let\next}
\setbox\strutbox=\hbox{\vrule height9.5pt depth4.5pt width0pt}
\global\newcount\ftno \global\ftno=0
\def\foot{\global\advance\ftno by1\footnote{$^{\the\ftno}$}}
%
\newwrite\ftfile   
\def\footend{\def\foot{\global\advance\ftno by1\chardef\wfile=\ftfile
$^{\the\ftno}$\ifnum\ftno=1\immediate\openout\ftfile=foots.tmp\fi%
\immediate\write\ftfile{\noexpand\smallskip%
\noexpand\item{f\the\ftno:\ }\pctsign}\findarg}%
\def\footatend{\vfill\eject\immediate\closeout\ftfile{\parindent=20pt
\centerline{\bf Footnotes}\nobreak\bigskip\input foots.tmp }}}
\def\footatend{}
%
%
\global\newcount\refno \global\refno=1
\newwrite\rfile
\def\ref{[\the\refno]\nref}
\def\nref#1{\xdef#1{[\the\refno]}\writedef{#1\leftbracket#1}%
\ifnum\refno=1\immediate\openout\rfile=refs.tmp\fi
\global\advance\refno by1\chardef\wfile=\rfile\immediate
\write\rfile{\noexpand\item{#1\ }\reflabeL{#1\hskip.31in}\pctsign}\findarg}
\def\findarg#1#{\begingroup\obeylines\newlinechar=`\^^M\pass@rg}
{\obeylines\gdef\pass@rg#1{\writ@line\relax #1^^M\hbox{}^^M}%
\gdef\writ@line#1^^M{\expandafter\toks0\expandafter{\striprel@x #1}%
\edef\next{\the\toks0}\ifx\next\em@rk\let\next=\endgroup\else\ifx\next\empty%
\else\immediate\write\wfile{\the\toks0}\fi\let\next=\writ@line\fi\next\relax}}
\def\striprel@x#1{} \def\em@rk{\hbox{}} 
\def\lref{\begingroup\obeylines\lr@f}
\def\lr@f#1#2{\gdef#1{\ref#1{#2}}\endgroup\unskip}

\def\addref#1{\immediate\write\rfile{\noexpand\item{}#1}} 
\def\startrefs#1{\immediate\openout\rfile=refs.tmp\refno=#1}
\def\xref{\expandafter\xr@f}\def\xr@f[#1]{#1}
\def\refs#1{\count255=1[\r@fs #1{\hbox{}}]}
\def\r@fs#1{\ifx\und@fined#1\message{reflabel \string#1 is undefined.}%
\nref#1{need to supply reference \string#1.}\fi%
\vphantom{\hphantom{#1}}\edef\next{#1}\ifx\next\em@rk\def\next{}%
\else\ifx\next#1\ifodd\count255\relax\xref#1\count255=0\fi%
\else#1\count255=1\fi\let\next=\r@fs\fi\next}
%

%
\newwrite\ffile\global\newcount\figno \global\figno=1
\def\fig{fig.~\the\figno\nfig}
\def\nfig#1{\xdef#1{fig.~\the\figno}%
\writedef{#1\leftbracket fig.\noexpand~\the\figno}%
\ifnum\figno=1\immediate\openout\ffile=figs.tmp\fi\chardef\wfile=\ffile%
\immediate\write\ffile{\noexpand\medskip\noexpand\item{Fig.\ \the\figno. }
\reflabeL{#1\hskip.55in}\pctsign}\global\advance\figno by1\findarg}
\def\xfig{\expandafter\xf@g}\def\xf@g fig.\penalty\@M\ {}
\def\figs#1{figs.~\f@gs #1{\hbox{}}}
\def\f@gs#1{\edef\next{#1}\ifx\next\em@rk\def\next{}\else
\ifx\next#1\xfig #1\else#1\fi\let\next=\f@gs\fi\next}
\newwrite\lfile
{\escapechar-1\xdef\pctsign{\string\%}\xdef\leftbracket{\string\{}
\xdef\rightbracket{\string\}}\xdef\numbersign{\string\#}}

\def\writestop{\def\writestoppt{\immediate\write\lfile{\string\pageno%
\the\pageno\string\startrefs\leftbracket\the\refno\rightbracket%
\string\def\string\secsym\leftbracket\secsym\rightbracket%
\string\secno\the\secno\string\meqno\the\meqno}\immediate\closeout\lfile}}
\def\writestoppt{}\def\writedef#1{}
\def\seclab#1{\xdef #1{\the\secno}\writedef{#1\leftbracket#1}\wrlabeL{#1=#1}}
\def\subseclab#1{\xdef #1{\secsym\the\subsecno}%
\writedef{#1\leftbracket#1}\wrlabeL{#1=#1}}
\newwrite\tfile \def\writetoca#1{}
\def\leaderfill{\leaders\hbox to 1em{\hss.\hss}\hfill}
\def\writetoc{\immediate\openout\tfile=toc.tmp 
   \def\writetoca##1{{\edef\next{\write\tfile{\noindent ##1 
   \string\leaderfill {\noexpand\number\pageno} \par}}\next}}}

%
\catcode`\@=12 
%
\edef\tfontsize{\ifx\answ\bigans scaled\magstep3\else scaled\magstep4\fi}
\font\titlerm=cmr10 \tfontsize \font\titlerms=cmr7 \tfontsize
\font\titlermss=cmr5 \tfontsize \font\titlei=cmmi10 \tfontsize
\font\titleis=cmmi7 \tfontsize \font\titleiss=cmmi5 \tfontsize
\font\titlesy=cmsy10 \tfontsize \font\titlesys=cmsy7 \tfontsize
\font\titlesyss=cmsy5 \tfontsize \font\titleit=cmti10 \tfontsize
\skewchar\titlei='177 \skewchar\titleis='177 \skewchar\titleiss='177
\skewchar\titlesy='60 \skewchar\titlesys='60 \skewchar\titlesyss='60
\def\titlefont{\def\rm{\fam0\titlerm}
\textfont0=\titlerm \scriptfont0=\titlerms \scriptscriptfont0=\titlermss
\textfont1=\titlei \scriptfont1=\titleis \scriptscriptfont1=\titleiss
\textfont2=\titlesy \scriptfont2=\titlesys \scriptscriptfont2=\titlesyss
\textfont\itfam=\titleit \def\it{\fam\itfam\titleit}\rm}
 \ifx\answ\bigans\else scaled\magstep1\fi
\ifx\answ\bigans\def\abstractfont{\tenpoint}\else
\font\abssl=cmsl10 scaled \magstep1
\font\absrm=cmr10 scaled\magstep1 \font\absrms=cmr7 scaled\magstep1
\font\absrmss=cmr5 scaled\magstep1 \font\absi=cmmi10 scaled\magstep1
\font\absis=cmmi7 scaled\magstep1 \font\absiss=cmmi5 scaled\magstep1
\font\abssy=cmsy10 scaled\magstep1 \font\abssys=cmsy7 scaled\magstep1
\font\abssyss=cmsy5 scaled\magstep1 \font\absbf=cmbx10 scaled\magstep1
\skewchar\absi='177 \skewchar\absis='177 \skewchar\absiss='177
\skewchar\abssy='60 \skewchar\abssys='60 \skewchar\abssyss='60
\def\abstractfont{\def\rm{\fam0\absrm}
\textfont0=\absrm \scriptfont0=\absrms \scriptscriptfont0=\absrmss
\textfont1=\absi \scriptfont1=\absis \scriptscriptfont1=\absiss
\textfont2=\abssy \scriptfont2=\abssys \scriptscriptfont2=\abssyss
\textfont\itfam=\bigit \def\it{\fam\itfam\bigit}\def\footnotefont{\tenpoint}%
\textfont\slfam=\abssl \def\sl{\fam\slfam\abssl}%
\textfont\bffam=\absbf \def\bf{\fam\bffam\absbf}\rm}\fi
\def\tenpoint{\def\rm{\fam0\tenrm}
\textfont0=\tenrm \scriptfont0=\sevenrm \scriptscriptfont0=\fiverm
\textfont1=\teni  \scriptfont1=\seveni  \scriptscriptfont1=\fivei
\textfont2=\tensy \scriptfont2=\sevensy \scriptscriptfont2=\fivesy
\textfont\itfam=\tenit \def\it{\fam\itfam\tenit}\def\footnotefont{\ninepoint}%
\textfont\bffam=\tenbf \def\bf{\fam\bffam\tenbf}\def\sl{\fam\slfam\tensl}\rm}
\font\ninerm=cmr9 \font\sixrm=cmr6 \font\ninei=cmmi9 \font\sixi=cmmi6 
\font\ninesy=cmsy9 \font\sixsy=cmsy6 \font\ninebf=cmbx9 
\font\nineit=cmti9 \font\ninesl=cmsl9 \skewchar\ninei='177
\skewchar\sixi='177 \skewchar\ninesy='60 \skewchar\sixsy='60 
\def\ninepoint{\def\rm{\fam0\ninerm}
\textfont0=\ninerm \scriptfont0=\sixrm \scriptscriptfont0=\fiverm
\textfont1=\ninei \scriptfont1=\sixi \scriptscriptfont1=\fivei
\textfont2=\ninesy \scriptfont2=\sixsy \scriptscriptfont2=\fivesy
\textfont\itfam=\ninei \def\it{\fam\itfam\nineit}\def\sl{\fam\slfam\ninesl}%
\textfont\bffam=\ninebf \def\bf{\fam\bffam\ninebf}\rm} 
%
%
\def\noblackbox{\overfullrule=0pt}
\hyphenation{anom-aly anom-alies coun-ter-term coun-ter-terms}
\def\inv{^{\raise.15ex\hbox{${\scriptscriptstyle -}$}\kern-.05em 1}}

\def\Dsl{\,\raise.15ex\hbox{/}\mkern-13.5mu D} 
\def\dsl{\raise.15ex\hbox{/}\kern-.57em\partial}

 \def\Tr{{\rm Tr}}
\font\bigit=cmti10 scaled \magstep1
\def\lspace{\ifx\answ\bigans{}\else\qquad\fi}
\def\lbspace{\ifx\answ\bigans{}\else\hskip-.2in\fi} 
\def\boxeqn#1{\vcenter{\vbox{\hrule\hbox{\vrule\kern3pt\vbox{\kern3pt
        \hbox{${\displaystyle #1}$}\kern3pt}\kern3pt\vrule}\hrule}}}
\def\mbox#1#2{\vcenter{\hrule \hbox{\vrule height#2in
                \kern#1in \vrule} \hrule}}  
%

\def\darr#1{\raise1.5ex\hbox{$\leftrightarrow$}\mkern-16.5mu #1}

\def\half{{\textstyle{1\over2}}} 
\def\roughly#1{\raise.3ex\hbox{$#1$\kern-.75em\lower1ex\hbox{$\sim$}}}

\openup -1pt
\input epsf
\expandafter\ifx\csname pre amssym.tex at\endcsname\relax \else\endinput\fi
\expandafter\chardef\csname pre amssym.tex at\endcsname=\the\catcode`\@
\catcode`\@=11
\ifx\undefined\newsymbol \else \begingroup\def\input#1 {\endgroup}\fi
\expandafter\ifx\csname amssym.def\endcsname\relax \else\endinput\fi
\expandafter\edef\csname amssym.def\endcsname{%
       \catcode`\noexpand\@=\the\catcode`\@\space}
\catcode`\@=11
\def\undefine#1{\let#1\undefined}
\def\newsymbol#1#2#3#4#5{\let\next@\relax
 \ifnum#2=\@ne\let\next@\msafam@\else
 \ifnum#2=\tw@\let\next@\msbfam@\fi\fi
 \mathchardef#1="#3\next@#4#5}
\def\mathhexbox@#1#2#3{\relax
 \ifmmode\mathpalette{}{\m@th\mathchar"#1#2#3}%
 \else\leavevmode\hbox{$\m@th\mathchar"#1#2#3$}\fi}
\def\hexnumber@#1{\ifcase#1 0\or 1\or 2\or 3\or 4\or 5\or 6\or 7\or 8\or
 9\or A\or B\or C\or D\or E\or F\fi}
\font\tenmsa=msam10
\font\sevenmsa=msam7
\font\fivemsa=msam5
\newfam\msafam
\textfont\msafam=\tenmsa
\scriptfont\msafam=\sevenmsa
\scriptscriptfont\msafam=\fivemsa
\edef\msafam@{\hexnumber@\msafam}
\mathchardef\dabar@"0\msafam@39
\def\maltese{{\mathhexbox@\msafam@7A}}
\font\tenmsb=msbm10
\font\sevenmsb=msbm7
\font\fivemsb=msbm5
\newfam\msbfam
\textfont\msbfam=\tenmsb
\scriptfont\msbfam=\sevenmsb
\scriptscriptfont\msbfam=\fivemsb
\edef\msbfam@{\hexnumber@\msbfam}
\def\Bbb#1{{\fam\msbfam\relax#1}}
\def\widehat#1{\setbox\z@\hbox{$\m@th#1$}%
 \ifdim\wd\z@>\tw@ em\mathaccent"0\msbfam@5B{#1}%
 \else\mathaccent"0362{#1}\fi}
\def\widetilde#1{\setbox\z@\hbox{$\m@th#1$}%
 \ifdim\wd\z@>\tw@ em\mathaccent"0\msbfam@5D{#1}%
 \else\mathaccent"0365{#1}\fi}
\font\teneufm=eufm10
\font\seveneufm=eufm7
\font\fiveeufm=eufm5
\newfam\eufmfam
\textfont\eufmfam=\teneufm
\scriptfont\eufmfam=\seveneufm
\scriptscriptfont\eufmfam=\fiveeufm

\csname amssym.def\endcsname
\relax
\newsymbol\smallsetminus 2272
\noblackbox
\newcount\figno
\figno=0
\def\mathrm#1{{\rm #1}}
\def\fig#1#2#3{
\par\begingroup\parindent=0pt\leftskip=1cm\rightskip=1cm\parindent=0pt
\baselineskip=11pt
\global\advance\figno by 1
\midinsert
\epsfxsize=#3
\centerline{\epsfbox{#2}}
\vskip 12pt
\centerline{{\bf Figure \the\figno} #1}\par
\endinsert\endgroup\par}
\font\tenmsb=msbm10       \font\sevenmsb=msbm7
\font\fivemsb=msbm5       \newfam\msbfam
\textfont\msbfam=\tenmsb  \scriptfont\msbfam=\sevenmsb
\scriptscriptfont\msbfam=\fivemsb
\def\Bbb#1{{\fam\msbfam\relax#1}}

\def\Zop{{\Bbb Z}}
\def\Cop{{\Bbb C}}
\def\Nop{{\Bbb N}}

\def\bbbone {{\mathchoice {\rm 1\mskip-4mu l} {\rm 1\mskip-4mu l}
{\rm 1\mskip-4.5mu l} {\rm 1\mskip-5mu l}}}
\def\bbbc{{\mathchoice {\setbox0=\hbox{$\displaystyle\rm C$}\hbox{\hbox
to0pt{\kern0.4\wd0\vrule height0.9\ht0\hss}\box0}}
{\setbox0=\hbox{$\textstyle\rm C$}\hbox{\hbox
to0pt{\kern0.4\wd0\vrule height0.9\ht0\hss}\box0}}
{\setbox0=\hbox{$\scriptstyle\rm C$}\hbox{\hbox
to0pt{\kern0.4\wd0\vrule height0.9\ht0\hss}\box0}}
{\setbox0=\hbox{$\scriptscriptstyle\rm C$}\hbox{\hbox
to0pt{\kern0.4\wd0\vrule height0.9\ht0\hss}\box0}}}}
\def\figlabel#1{\xdef#1{\the\figno}}
\def\pano{\par\noindent}

\def\noi{\noindent}
\def\pmb#1{\setbox0=\hbox{#1}%
 \kern-.025em\copy0\kern-\wd0
 \kern.05em\copy0\kern-\wd0
 \kern-.025em\raise.0433em\box0 }

\def\half{{1\over 2}}



\def\H{{\cal H}}

\def\A{{\cal{A}}}
\def\Tr{{\hbox{Tr}}}

\def\su{\hbox{$\widehat{\rm su}$}}
\def\smallmatrix{\raise1.6pt\hbox{${\scriptscriptstyle 
{{\;0\, \; 1} \choose {-\!1\;0}} 
}$}}
\def\smallmatrixone{\raise1.6pt\hbox{${\scriptscriptstyle 
{{1\,\;0} \choose {\,0\; \, 1}} 
}$}}
\def\hsmallsetminus{\hbox{\raise1.5pt\hbox{$\smallsetminus$}}}
\def\tilM{\hbox{${\scriptstyle \widetilde{\phantom M}}$}\hskip-9pt
   \raise1.3pt\hbox{${\scriptstyle M}$}\,}
\def\ie{{\it i.e.}}
\def\bq{{\bar{q}}}
\def\sn{\smallskip\noindent}
\def\mn{\medskip\noindent}


\def\figin{\epsfcheck\figin}\def\figins{\epsfcheck\figins}
\def\epsfcheck{\ifx\epsfbox\UnDeFiNeD
\message{(NO epsf.tex, FIGURES WILL BE IGNORED)}
\gdef\figin##1{\vskip2in}\gdef\figins##1{\hskip.5in}
\else\message{(FIGURES WILL BE INCLUDED)}%
\gdef\figin##1{##1}\gdef\figins##1{##1}\fi}
\def\DefWarn#1{}
\def\figinsert{\goodbreak\midinsert}
\def\ifig#1#2#3{\DefWarn#1\xdef#1{fig.~\the\figno}
\writedef{#1\leftbracket fig.\noexpand~\the\figno}%
\figinsert\figin{\centerline{#3}}\medskip\centerline{\vbox{\baselineskip12pt
\advance\hsize by -1truein\noindent\footnotefont{\bf Fig.~\the\figno:} #2}}
\bigskip\endinsert\global\advance\figno by1}

\def\gbs{|\!|g\rangle\!\rangle}
\def\sgbs{|\!|g,\eta\,\rangle\!\rangle}
\def\ishjmn{|\,j;m,n\,\rangle\!\rangle}
\def\braishjmn{\langle\!\langle\,j;m,n\,|}

\lref\rsone{A.\ Recknagel, V.\ Schomerus, {\it Boundary deformation
theory and moduli spaces of D-branes}, Nucl.\ Phys.\ {\bf B545}, 233 
(1999); {\tt hep-th/9811237}.}

\lref\klebpol{I.R.\ Klebanov, A.M.\ Polyakov, {\it Interaction of
discrete states in two-dimensional string theory},
Mod.\ Phys.\ Lett.\ {\bf A6}, 3273 (1991); {\tt hep-th/9109032}.} 

\lref\cklm{C.G.\ Callan, I.R.\ Klebanov, A.W.\ Ludwig, J.M.\
Maldacena, {\it Exact solution of a boundary conformal field theory}, 
Nucl.\ Phys.\ {\bf B422}, 417 (1994); {\tt hep-th/9402113}.}

\lref\polthor{J.\ Polchinski, L.\ Thorlacius, {\it Free fermion
representation of a boundary conformal field theory}, Phys.\ Rev.\ 
{\bf D50}, 622 (1994); {\tt hep-th/9404008}.}

\lref\fk{I.B.\ Frenkel, V.G.\ Kac, {\it Basic representations of
affine Lie algebras and dual resonance models}, Invent.\ Math.\ 
{\bf 62}, 23 (1981).} 

\lref\segal{G.B.\ Segal, {\it Unitary representations of some infinite 
dimensional groups}, Commun.\ Math.\ Phys.\ {\bf 80}, 301 (1981).} 

\lref\ham{M.\ Hamermesh, {\it Group theory and its applications to
physical problems}, Addison-Wesley (1962).}

\lref\cardy{J.L.\ Cardy, {\it Boundary conditions, fusion rules and
the Verlinde formula}, Nucl.\ Phys.\ {\bf B324}, 581 (1989).} 

\lref\DGH{L.\ Dixon, P.\ Ginsparg, J.\ Harvey, {\it $\hat c=1$
superconformal field theory}, Nucl.\ Phys.\ {\bf B306}, 470 (1988).}

\lref\bg{O.\ Bergman, M.R.\ Gaberdiel, {\it A non-supersymmetric open 
string theory and S-duality}, Nucl.\ Phys.\ {\bf B499}, 183 (1997); 
{\tt hep-th/9701137}.}

\lref\feiginfuchs{B.L.\ Feigin, D.B.\ Fuchs, {\it Invariant
skew-symmetric differential operators on the line and Verma modules
over the Virasoro algebra}, Funct. Anal. Appl. {\bf 16}, 114 (1982); 
\quad {\it Verma modules over the Virasoro algebra}, in:\ Lecture 
Notes in Mathematics {\bf 1060}, Springer 1984.}

\lref\grwone{M.R.\ Gaberdiel, A.\ Recknagel, G.M.T.\ Watts, 
{\it The conformal boundary states for SU(2) at level $1$}; 
{\tt hep-th/0108102}.} 

\lref\knapp{A.W.\ Knapp, {\it Representation theory of semisimple
groups: an overview based on examples}, Princeton University Press,
Princeton (1986).}

\lref\cohnfried{J.D.\ Cohn, D.\ Friedan, {\it Super characters and 
chiral asymmetry in superconformal field theory}, Nucl.\ Phys.\ 
{\bf B296}, 779 (1988).}

\lref\DixHar{L.J.\ Dixon, J.A.\ Harvey, {\it String theories
in ten dimensions without spacetime supersymmetry}, Nucl.\ Phys.\
{\bf B274}, 93 (1986).}

\lref\seiwitt{N.\ Seiberg, E.\ Witten, {\it Spin structures
in string theory}, Nucl.\ Phys.\ {\bf B276}, 272 (1986).}

\lref\sen{A.\ Sen, {\it BPS D-branes on non-supersymmetric
cycles}, J.\ High Energy Phys.\  {\bf 9812}, 021 (1998); 
{\tt hep-th/9812031}.}

\lref\arfken{G.\ Arfken, {\it Mathematical methods for physicists},
Academic Press (1970).}

\lref\friedan{D.\ Friedan, {\it The space of conformal boundary
conditions for the $c=1$ Gaussian model}, unpublished note (1999).} 

\lref\gutperle{M.\ Gutperle, {\it Non-BPS D-branes and enhanced
symmetry in an asymmetric orbifold}, J.\ High Energy Phys.\ {\bf 0008}, 
036 (2000); {\tt hep-th/0007126}.}

\lref\romek{R.A.\ Janik, {\it Exceptional boundary states at $c=1$}, 
to appear.}

\lref\thompson{D.M.\ Thompson, {\it Descent relations in type 0A/0B}, 
{\tt hep-th/0105314}.}

\lref\mms{J.\ Maldacena, G.\ Moore, N.\ Seiberg, {\it  Geometrical 
interpretation of D-branes in gauged WZW models}, 
{\tt hep-th/0105038}.} 

\lref\fsb{L.\ Birke, J.\ Fuchs, C.\ Schweigert, {\it Symmetry breaking 
boundary conditions and WZW orbifolds}, Adv.\ Theor.\ Math.\ Phys.\ 
{\bf 3}, 671 (1999); {\tt hep-th/9905038}.}

\lref\pol{J.\ Polchinski, {\it Combinatorics of boundaries in string
theory}, Phys.\ Rev.\ {\bf D50}, 6041 (1994); {\tt hep-th/9407031}.}

\lref\FS{J.\ Fuchs, C.\ Schweigert, {\sl A classifying algebra for
boundary conditions}, Phys.\ Lett.\ {\bf B414}, 251 (1997); 
{\tt hep-th/9708141}.} 

\lref\PSS{G.\ Pradisi, A.\ Sagnotti, Y.S.\ Stanev, {\it Completeness   
conditions for boundary operators in 2d conformal field theory}, 
Phys.\ Lett.\ {\bf B381}, 97 (1996); {\tt hep-th/9603097}.}

\lref\BPPZ{R.E.\ Behrend, P.A.\ Pearce, V.B.\ Petkova, J.-B.\ Zuber,  
{\it Boundary conditions in rational conformal field theories}, 
Nucl.\ Phys.\ {\bf B579}, 707 (2000); {\tt hep-th/9908036}.}

\lref\Lew{D.C.\ Lewellen, {\it Sewing constraints for conformal field 
theories on surfaces with boundaries}, Nucl.\ Phys.\ {\bf B372}, 654
(1992).} 

\lref\GG{M.B.\ Green, M.\ Gutperle, {\it Symmetry breaking at 
enhanced symmetry points}, Nucl.\ Phys.\ {\bf B460}, 77 (1996); 
{\tt hep-th/9509171}.}

\Title{\vbox{
\hbox{hep--th/0108238}
\hbox{KCL-MTH-01-38}}}
{\vbox{\centerline{
Conformal boundary states for free bosons}
\vskip16pt
\centerline{and fermions}
}}
\centerline{M.R.\ Gaberdiel\footnote{}{\hskip-12pt{\tt e-mails:
mrg@mth.kcl.ac.uk, anderl@mth.kcl.ac.uk}}$\;\,$ 
and$\;\,$ A.\ Recknagel
}
\bigskip
\centerline{\it Department of Mathematics, King's College London}
\centerline{\it Strand, London WC2R 2LS, U.K.}
\smallskip
\vskip2cm
\centerline{\bf Abstract}
\bigskip

{\narrower \noindent
A family of conformal boundary states for a free boson
on a circle is constructed. The family contains superpositions of 
conventional U(1)-preserving Neumann and Dirichlet branes, but for 
general parameter values the boundary states are fundamental and 
preserve only the conformal symmetry. The relative overlaps satisfy 
Cardy's condition, and each boundary state obeys the factorisation
constraint. It is also argued that, together with the conventional 
Neumann and Dirichlet branes, these boundary states already account
for all fundamental conformal D-branes of the free boson theory. The
results can be generalised to  the situation with $N=1$ world-sheet
supersymmetry, for which the family of boundary states interpolates
between superpositions of non-BPS branes and combinations of
conventional brane anti-brane pairs. 

} 
\bigskip 

\Date{\hbox{\quad\tt August 2001}}

\newsec{Introduction}

\noi
Most D-branes that have been constructed in string theory 
are quite special in that they preserve some large symmetry
algebra. For example, the standard Dirichlet or Neumann branes of
type IIA or IIB string theory leave all the U(1) current symmetries 
(associated to the different coordinates in spacetime)
invariant. This is to say, the relevant boundary states satisfy the
boundary conditions
\eqn\introone{
\left(\alpha^\mu_{n} \pm \overline{\alpha}^\mu_{-n} \right)
\,|\!|B\rangle\!\rangle=0\,,}
where $\alpha^\mu_n$ are the modes associated to $X^\mu$, and where 
the sign determines whether  $X^\mu$ obeys a Dirichlet or a
Neumann condition on the world-sheet boundary. However, as was already
remarked early on in \refs{\pol}, the most general branes that should be 
relevant for (bosonic) string theory only preserve conformal
invariance, \ie\ they only satisfy
\eqn\introtwo{
\bigl(\, L_n - \overline{L}_{-n}\,\bigr)\,|\!|B\rangle\!\rangle
=0\,.}
If a boundary state $|\!|B\rangle\!\rangle$ satisfies \introone\ 
for all $\mu$, then \introtwo\  automatically holds, since the
Virasoro modes are bilinears in the current modes. However, in
general, \introtwo\ does not imply \introone. It is therefore
important to understand and characterise the more general class 
of boundary states that only satisfy \introtwo. 

In this paper we shall analyse this question in detail for the 
simplest example, a free boson on a circle.\footnote{$^\star$}{For a 
different class of theories the similar problem of finding boundary
states with reduced symmetry was recently analysed in \refs{\mms}.} We
shall also consider the supersymmetric analogue, where we have a free
boson and a free fermion. In the bosonic case we shall mainly
concentrate on the situation when the radius is a fraction 
${M\over N}$ of the self-dual radius (where $M$ and $N$ are coprime),
for which we can give an explicit construction of a family of
conformal boundary states parametrised by the manifold
SU(2)$/\Zop_M\times\Zop_N$. (The action of the cyclic groups is
explained in Section~4.) These boundary states have the property  
that all relative overlaps (boundary partition functions on the strip)
satisfy Cardy's condition. We can also provide strong arguments 
that, together with the standard Dirichlet and Neumann brane states, 
these boundary states are already all fundamental conformal boundary
states. Some of these results had been anticipated by Friedan in
unpublished  work \refs{\friedan}.

The parameter space SU(2)$/\Zop_M\times\Zop_N$ contains two special 
submanifolds, one describing $M$ Dirichlet branes placed at equidistant 
circle points $x+{\sqrt{2}\pi k\over N}$ with $k=1,\ldots,M$, while the 
other submanifold corresponds to superpositions of $N$ Neumann 
branes with evenly spaced Wilson lines. The general D-branes in the 
family interpolate
between these two extremal configurations. In fact, the Dirichlet or
Neumann brane configurations merge into intermediate boundary states
that can no longer be thought of as  superpositions of fundamental
branes. These intermediate branes are themselves fundamental, and 
they do not preserve the U(1) symmetry.  

In the case with world-sheet supersymmetry, the essential features are
the same, although there are a number of minor differences. Most
notably, if we want to preserve world-sheet supersymmetry, the boundary
states should now also preserve the supercurrent. In addition, in
order to obtain a modular invariant partition function for the closed
string sector, we have to impose a GSO projection, and this gives rise
to additional constraints for the boundary states. The moduli space of
superconformal boundary states has essentially the same structure as
before, but now the special submanifolds correspond to collections of
(unstable) non-BPS branes on the one extreme and to brane anti-brane  
combinations on the other (the latter are Dirichlet or Neumann 
depending on the radius $R=M/N$). As before, the general boundary 
states in the family interpolate between these configurations. In
addition, the moduli space contains an extra branch that describes
unstable non-BPS branes similar to those discovered by Sen.
\sn
Neither the bosonic nor the fermionic theory are `rational' with
respect to the current or supersymmetric current algebra. The D-branes
we consider do not even preserve this algebra, but only the Virasoro
or super Virasoro subalgebra; thus they are `symmetry breaking
boundary conditions'. However, in contrast to the framework outlined in
\refs{\fsb}, our theory is truly non-rational. As a consequence, the
usual techniques that are employed for rational theories (in
particular, Cardy's construction \refs{\cardy}) are not available 
in our case. In addition, the question of completeness of the 
boundary states is not a purely algebraic problem any more, but 
involves analytic considerations. That one can nevertheless obtain 
explicit results rests essentially on the fact that irreducible 
Virasoro representations at $c=1$ are known in great detail. 
\mn
The paper is organised as follows. In Section 2 we fix our conventions
and describe the (conformal) Ishibashi states of the theory. In
Section~3 we revisit and reformulate the construction of boundary 
states for the self-dual radius, and we determine the various overlaps
explicitly. These techniques are then generalised in Section~4 to the
case where the radius is an arbitrary fraction of the self-dual
radius. Section~5 deals with the zero radius limit in which we recover
some results of \refs{\polthor}, which had been obtained using
completely different methods. In Section~6 we analyse the
`factorisation constraints' and show that there are essentially no
other fundamental conformal branes. We also explain in some detail in
which sense  these boundary states span the whole space of conformal
boundary conditions. Section~7 deals with the modifications that arise  
in the case with world-sheet supersymmetry, and Section~8 contains
some comments regarding the stability of the corresponding D-branes in
a  ten-dimensional setting. Finally, we end in Section~9 with some
open problems.

\newsec{Notations, conventions and some basic results}

\noi
We want to determine a class of conformal boundary states for a 
free boson of conformal charge $c=1$, and later for the 
generalisation to the $N=1$ superconformal case.
We shall mainly be interested in the situation where the boson is
compactified on a circle of radius $R$ (although we shall also make
some statements about the limit $R\rightarrow 0$). Choosing 
the convention $\alpha'=\half$, the left- and right-moving
momenta are given as 
\eqn\momenta{ 
\left(p_L,p_R\right) = \left( {\hat{m}\over 2R} + \hat{n} R, 
{\hat{m}\over 2R} - \hat{n} R \right) }
with $\hat{m},\hat{n}\in\Zop$. The partition function of the theory
is then 
\eqn\part{
Z(q,\bq) = \sum_{\hat{m},\hat{n}\in\Zop} 
\vartheta_{p_L(\hat{m},\hat{n})}(q) \,
\vartheta_{p_R(\hat{m},\hat{n})}(\bq) \,,}
where we use the slightly non-standard notation 
\eqn\th{
\vartheta_s(q) = {q^{\half s^2} \over \eta(q)}}
and where $\eta(q)$ is the usual Dedekind $\eta$-function,
\eqn\ett{
\eta(q) = q^{1\over 24} \prod_{n=1}^{\infty} (1-q^n)\,.}
In these units, the self-dual radius is 
$R_{\rm s.d.}={1\over\sqrt{2}}$; at this radius, the theory is
actually equivalent to a level $k=1$ affine $\su(2)$ theory. (This is
the content of the vertex operator construction of Frenkel, Kac and
Segal \refs{\fk,\segal}.) 

In the following it will be important to understand the decomposition
of the above U(1) representations in terms of the Virasoro
algebra. At $c=1$, every highest weight representation of the Virasoro
algebra is irreducible except if the conformal weight is of the form 
$h=j^2$ for some $j=0,\half,1,\ldots$;
in this case, the Virasoro representation has a single null-vector at
level $2j+1$ \refs{\feiginfuchs}.\footnote{$^\dagger$}{All of these
representations are actually representations of the chiral $c=1$
Virasoro theory (\ie\ the chiral vacuum sector or the vertex operator
algebra). Since there is a continuum of such representations, the
conformal field theory is not `rational'. Thus, Cardy's technique
\refs{\cardy} for finding boundary states does not directly apply.}
The corresponding characters are then given by  
\eqn\Vir{\eqalign{
h\ne j^2 \qquad & \chi_h(q) = {q^h \over \eta(q)}\;, \cr
h = j^2  \qquad & \chi_h(q) = \vartheta_{\sqrt{2}j }(q) 
                               - \vartheta_{\sqrt{2}(j+1)}(q)\;.}}
If $p_L\ne \sqrt{2}\,j$ for any $j\in\half\Zop$, then the irreducible
U(1)-representation with highest weight $p_L$ also defines an
irreducible Virasoro representation. On the other hand, if  
$p_L= \sqrt{2}\,j$ for some $j\in\half\Zop$ we have the decomposition
\eqn\decomp{
\H^{U(1)}_{j} = \bigoplus_{l=0}^{\infty}\, \H^{\rm Vir}_{|j|+l} \,,}
where $\H^{\rm Vir}_{k}$ (with $k\geq 0$) is the irreducible Virasoro
representation with highest weight $h=k^2$. (This representation can
be obtained by dividing out the subrepresentation generated by the
null-vector at level $2k+1$ from the Verma module over the highest 
weight state.) Here we have used the (obvious) character identity 
\eqn\iden{
\vartheta_{\sqrt{2} j } = \sum_{l=0}^{\infty} 
   \left(\vartheta_{\sqrt{2}(j+l)} -
           \vartheta_{\sqrt{2}(j+l+1)}\right) \,.}
Analogous statements obviously hold for the right-moving
representations. 

We are only interested in boundary states that preserve the conformal
symmetry (so that the corresponding field theory, which is defined on a 
surface with a boundary, is again conformal). In order to describe the
most general such boundary state, we need to rewrite the spectrum
of the closed string theory (\ie\ the spectrum of the conformal
field theory defined on the sphere) in terms of Virasoro
representations. For each sector for which the left- and right-moving
Virasoro representation are equivalent, we can construct an Ishibashi
state; the most general boundary state is then a linear combination of
these Ishibashi states, with coefficients constrained by a number of 
consistency conditions.

In this paper we shall construct a class of boundary states that are
built up from Ishibashi states associated with {\it degenerate} 
representations of the Virasoro algebra, \ie\ with highest weights 
$h=j^2$ for some $j\in\half\Zop$. The set of representations that 
are actually present in the closed string spectrum (as well as the 
multiplicity with which they show up) depends critically on the value 
of $R$. Degenerate Virasoro representations (other than those in 
the U(1) vacuum sector) only occur when $R$ is a fraction of the 
self-dual radius,
\eqn\Rvals{
R= {M\over N} {1\over\sqrt{2}} \,,}
where $M$ and $N$ are coprime positive integers. 
In order to describe the set of Ishibashi states for a general radius 
of the above form, it is convenient to look at the self-dual case 
$M=N=1$ first (see also \refs{\cklm,\polthor,\rsone}). In this case, 
the momenta are of the form 
\eqn\self{
(p_L,p_R) = {1\over \sqrt{2}} (\hat{m}+\hat{n},\hat{m}-\hat{n}) \,,}
and the corresponding Virasoro highest weights are given by 
\eqn\highest{
h = \half p_L^2 = \left({\hat{m} + \hat{n} \over 2}\right)^2 
= m^2\ , \qquad
\bar{h} = \half p_R^2 = \left({\hat{m} - \hat{n} \over 2}\right)^2 
= n^2\,,}
where we have defined 
\eqn\defmn{
m = {\hat{m} + \hat{n} \over 2} \qquad
n = {\hat{m} - \hat{n} \over 2} \,.}
By construction, $m$ and $n$ are half-integers whose difference is 
always integer. Because of \highest, all the U(1) representations 
that occur at the self-dual point give rise to degenerate Virasoro 
representations. It then follows from \decomp\ that for fixed $(m,n)$, 
the U(1)$\,\times\,$U(1) representation (whose highest weight states 
have momenta $(p_L,p_R)$) contains the Virasoro representation
$\H^{\rm Vir}_{j}\otimes \overline{\H}{}^{\rm Vir}_{j}$ provided that
both $|m|$ and $|n|$ are less or equal than $j$ and that $j-m$ (and
$j-n$) is integer. Conversely, if $(m,n)$ satisfy these constraints,
the representation 
$\H^{\rm Vir}_{j}\otimes\overline{\H}{}^{\rm Vir}_{j}$ appears 
precisely once in the corresponding U(1)$\,\times\,$U(1)
representation. Thus the possible Ishibashi states can be labelled by
the triples $(j;m,n)$ where $j$ is a non-negative half-integer, 
$m=-j,-j+1,\ldots,j$, and similarly for $n$. These triples can be 
thought of as labelling the possible matrix elements of the SU(2) 
representation $j$.  

For the more general case \Rvals, left- and right-moving momenta are 
of the form
\eqn\momgen{
\left(p_L,p_R\right) = {1\over \sqrt{2}} \left(  {\hat{n} N \over M} + 
{\hat{m} M \over N}, {\hat{n} N \over M} -{\hat{m} M \over N}\right)
=   \sqrt{2} \left( m , n \right) \,,} 
where $m$ and $n$ are defined in analogy to \defmn, and thus the 
conformal weights are 
\eqn\confwgen{
h = {1\over 4}\left({\hat{n} N \over M}+{\hat{m} M \over N}\right)^2 
  = m^2
\qquad
\bar{h} = 
{1\over 4}\left({\hat{n} N \over M}-{\hat{m} M \over N}\right)^2 
= n^2\,.}
In general, $m$ and $n$ are not half-integers any more, and therefore
also Ishibashi states for non-degenerate Virasoro representations
occur. If we concentrate on the degenerate Virasoro representations,
\ie\ if we choose $\hat{m},\hat{n}$ so that $m$ and $n$ are
half-integers, then $m-n$ and $m+n$ are necessarily integers. (This
follows from the fact that $(m-n)(m+n)=\hat{m}\hat{n}$.) Furthermore,
since $M$ and $N$ are coprime, we have the constraints
\eqn\constraints{
m+n = l N \ ,\quad m-n = l' M \qquad {\rm for\ some}\ \ l,l'\in\Zop}
on the $(m,n)$ that correspond to degenerate Virasoro representations. 
Conversely, if these constraints are satisfied, the spectrum contains
a Virasoro Ishibashi state corresponding to $j$ for every $j$ that is 
greater or equal to $|m|$ and $|n|$  and satisfies $j-m\in\Zop$. Thus 
the (degenerate) Ishibashi states that exist in the general case can 
be labelled by the triples $(j;m,n)$ as before where now $m$ and $n$ 
satisfy in addition \constraints; this characterisation of the Ishibashi 
states will be crucial in what follows.

\newsec{The boundary states at the self-dual radius}

\noi
As a warm-up, let us first analyse the boundary states for the theory
at the self-dual radius. Explicit expressions were for example given
in  \refs{\GG,\rsone}, where it was observed that there exists a
moduli space of boundary conditions that are connected by truly
marginal boundary fields and that can be parametrised by 
$g \in {\rm SU(2)}$. All the boundary states $\gbs$ in this
$S^3$-family preserve $\su(2)_1$, with gluing conditions 
\eqn\glue{
   \left( 
   {\rm Ad}_{(\,g \cdot\iota\,)}(J^a_m)  + J^a_{-m} 
     \right) \gbs  = 0 
\qquad {\rm where}\quad\  
  \iota = \pmatrix{0 & 1 \cr -1 & 0}
}
and for all $m \in \Zop$. 
In the following we shall concentrate on checking
the Cardy condition (which was only implicitly solved in \rsone). In
particular, we shall obtain a compact formula for the open string
spectrum between arbitrary boundary states. A modification of the
technique we shall use in the derivation of this result can then be
used to solve the Cardy condition for all radii of the form
\Rvals. 

\noindent For each group element $g\in {\rm SU(2)}$, we define the
boundary state 
\eqn\boundsu{
\gbs = {1\over 2^{1\over 4}}
\sum_{j,m,n} D^j_{m,n}(g)\, \ishjmn \,,}
where $D^j_{m,n}(g)$ is the matrix element of $g$ in the
representation $j$, and $\ishjmn$ denotes the
Virasoro Ishibashi state labelled by the triple $(j;m,n)$ as above. 
The gluing condition fixes the Ishibashi state up to an overall
factor; we shall always  assume that they are normalised so that    
\eqn\norma{
\braishjmn\, q^{\half(L_0+\bar{L}_0-{c\over12})}\, 
\ishjmn = \chi_{j^2} (q) \,,}
where $\chi_{j^2}$ denotes the character of the Virasoro highest
weight representation with $h=j^2$ (see \Vir\ above). An explicit
formula for the matrix elements is given by
\eqn\matrixelem{\eqalign{
D^j_{m,n}(g) & = \sum_{l=\max(0,n-m)}^{\min(j-m,j+n)}
{\left[ (j+m)!\, (j-m)!\, (j+n)!\, (j-n)! \right]^{\half} \over
(j-m-l)!\, (j+n-l)!\, l!\, (m-n+l)!} \cr
& \hskip120pt \times a^{j+n-l} (a^*)^{j-m-l} b^{m-n+l} (-b^*)^{l}\,,}}
where we have written $g\in {\rm SU(2)}$ as 
\eqn\group{
g= \pmatrix{ a & b \cr - b^\ast & a^\ast} \,.}
This differs slightly from the formula given in \refs{\ham}; with the
definition \matrixelem\ we then have 
$D^j_{m,n}(g\cdot h) = \sum_l D^j_{m,l}(g)\, D^j_{l,n}(h)$.

If we set $g=e$ in \boundsu, the boundary state describes the standard
Dirichlet brane (at $x=0$). On the other hand, the standard Neumann
brane (without any Wilson line) corresponds to the choice $a=0,b=1$ in
\group. The prefactor of \boundsu\ is the familiar factor 
$(2R)^{-{1\over 2}}$ that arises in the normalisation of conventional
Dirichlet boundary states.  

One of the consistency conditions that have to be satisfied by the
boundary states is the constraint --- usually referred to as the Cardy
condition \refs{\cardy}--- that the overlap of any two boundary
states gives rise, upon modular transformation, to an open string
one-loop amplitude. In particular, this requires that the different
`open string representations' contributing to the one-loop
amplitude occur with positive integer multiplicities. We shall now
demonstrate explicitly that the boundary states \boundsu\ satisfy this
constraint. In order to do so, we begin by rewriting the closed string
tree-level amplitude  
\eqn\calc{
\A= \langle\!\langle g_1 |\!|\, q^{\half(L_0+\bar{L}_0-{c\over12})}\,
   |\!| g_2 \rangle\!\rangle = 
{1\over \sqrt{2}} 
\sum_{j\in\half\Zop_+} \sum_{m,n} \left(D^j_{m,n}(g_1)\right)^\ast
D^j_{m,n}(g_2)\, \chi_{j^2}(q) \,.}
Since the representation defined by $j$ is unitary, we have 
\eqn\calcone{
\left(D^j_{m,n}(g_1)\right)^\ast = D^j_{n,m}(g_1^{-1})\,.}
For each fixed $j$, we can then perform the sum over $m$ and $n$, and
we obtain
\eqn\calctwo{
\sum_{m,n} D^j_{n,m}(g_1^{-1}) D^j_{m,n}(g_2) = 
\sum_n D^j_{n,n}(g_1^{-1} g_2) = \Tr_{j} (g_1^{-1} g_2) \,,}
where $\Tr_{j}$ is the trace in the $j$-th representation of
SU(2). Since the trace is invariant under conjugation, we can rotate 
$g_1^{-1} g_2$ to lie in the maximal torus of SU(2); then 
\eqn\calcthree{
\Tr_{j} (g_1^{-1} g_2) = {\sin((2j+1)\alpha) \over \sin\alpha}}
for some $\alpha\in[0,2\pi)$ that depends on $g_1^{-1} g_2$. (The
simplest way to determine $\alpha$ explicitly is to evaluate the
left hand side of \calcthree\ in the fundamental ($j=\half$)
representation of SU(2), where the right hand side is simply
$2\cos\alpha$.)

Next, we rewrite the Virasoro character $\chi_{j^2}$ in terms of the
$\vartheta$-functions as in \Vir, and thus obtain
\eqn\calcfour{\eqalign{
\sqrt{2} \A & = \sum_{j\in\half\Zop_+} 
{\sin((2j+1)\alpha) \over \sin\alpha} 
\left(\vartheta_{\sqrt{2}j}(q) - \vartheta_{\sqrt{2}(j+1)}(q) \right) \cr
& = \vartheta_0 (q) + \sum_{j=1}^{\infty} \vartheta_{\sqrt{2}j}(q)
\left({\sin((2j+1)\alpha) \over \sin\alpha} - 
{\sin((2j-1)\alpha) \over \sin\alpha} \right) \cr
& \quad + 2 \cos\alpha\cdot \vartheta_{1\over\sqrt{2}} (q) 
+ \sum_{j={3\over 2}}^{\infty} \vartheta_{\sqrt{2}j}(q)
\left({\sin((2j+1)\alpha) \over \sin\alpha} - 
{\sin((2j-1)\alpha) \over \sin\alpha} \right) \cr
& = \sum_{j\in\half\Zop} \cos(2j\alpha)\, \vartheta_{\sqrt{2}j}(q) \,,}}
where the first (second) sum in the middle equation
runs over integers (half-odd integers) only. Moreover, we have used
the identity   
\eqn\trigo{
{\sin((2j+1)\alpha) \over \sin\alpha} - 
{\sin((2j-1)\alpha) \over \sin\alpha} = 2 \cos(2j\alpha) \,,}
as well as the fact that $\vartheta_s(q)$ only depends on $|s|$. Under
a modular transformation, the $\vartheta$-functions \th\ behave as  
\eqn\modular{
\vartheta_s(q) = \int_{-\infty}^{\infty} dt\, e^{2\pi i t s} \,
       \vartheta_t(\tilde{q})\,,}
where $q=e^{2\pi i\tau}$ with $\Im\tau>0$ and 
$\tilde{q}=  e^{-{2\pi i \over \tau}}$. Putting these results
together, we then find that the overlap $\A = Z_{g_1g_2}(\tilde q)$
becomes, in the open string description, 
\eqn\calcfive{\eqalign{
Z_{g_1g_2}(\tilde q) & = {1\over \sqrt{2}} 
\int_{-\infty}^{\infty} dt\, \vartheta_t(\tilde{q})
\sum_{j\in\half\Zop} e^{2\pi i \sqrt{2} j t} \cos(2j\alpha)  \cr
& = {1\over \sqrt{2}} 
\int_{-\infty}^{\infty} dt\, \vartheta_t(\tilde{q})
\sum_{l\in\Zop} e^{i l (\sqrt{2}\pi t + \alpha)} \cr
& = {2 \pi\over \sqrt{2}} 
\int_{-\infty}^{\infty} dt\, \vartheta_t(\tilde{q})
\sum_{n\in\Zop} \delta(\sqrt{2}\pi t + \alpha + 2 \pi n) \cr
& = \sum_{n\in\Zop} 
\vartheta_{-{\alpha\over \sqrt{2}\pi} + \sqrt{2}n}(\tilde{q})\,.}}
Irrespective of the value of $\alpha$, this defines a positive integer
linear combination of Virasoro or U(1) characters. For $g_1=g_2$, the 
spectrum displays the full $\su(2)_1$-symmetry as expected. 

Since $\alpha$ only depends on $g_1^{-1} g_2$, the open string
spectrum between any of the above Dirichlet branes (labelled by 
$g\in {\rm SU(2)}$) and itself is the same for all $g$. This
agrees with the results found in \refs{\cklm,\polthor,\rsone}. 
However, \calcfive\ also determines the open string spectrum 
between two D-branes associated to different SU(2) elements.

\newsec{The generalisation to fractional radii}

\noi
We shall now show that, for any `fractional radius' of the form 
$R={M\over N}R_{\rm s.d.}$ as in \Rvals, there is again a conformal 
boundary state for each group element $g\in {\rm SU(2)}$. These
boundary states are given by formula \boundsu\ from above, except for
the  following two changes: (i) the $m$- and $n$-summations run only
over values which satisfy in addition \constraints, and (ii) the
overall normalisation is $\sqrt{MN}$ times larger than the one in
\boundsu. Since for $R \neq R_{\rm s.d.}$, the
Virasoro  Ishibashi states do not align into U(1) Ishibashi states,
the new boundary states will in general break the symmetry from U(1)
to Vir. If $M=1$, these states can be obtained by marginal
boundary deformations of an ordinary Dirichlet boundary state 
\refs{\cklm,\polthor,\rsone}, and the analogous statement holds 
for the $T$-dual situation. For other radii, they are not connected 
to a single U(1)-preserving Dirichlet boundary state by marginal 
deformations; however, as we shall see below, they are connected to
certain superpositions of ordinary Dirichlet branes.

In order to show that the corresponding open string spectrum still
satisfies the desired integrality property, we want to perform a
calculation similar to what we have done in the previous section. 
However, there is one new difficulty, namely that \calctwo\ is not 
applicable as it stands since we are not summing over {\it all}
values of $m$ and $n$ any more. We can circumvent this problem if 
we use the following trick. Let us introduce the group element
\eqn\trick{
\Gamma_M = \pmatrix{ e^{{\pi i \over M}} & 0 \cr 
0 & e^{-{\pi i \over M}}} \in {\rm SU(2)} \,.}
Using the explicit form of \matrixelem\ we find that 
\eqn\calcp{
D^j_{m,n} \left(\Gamma_M\right) 
= e^{{2\pi i \over M} m} \delta_{m,n} \,.}
Thus, 
\eqn\calcpone{
D^j_{m,n} \left(\Gamma_M^l \, g \, \Gamma_M^{-l}\right) = 
e^{{2\pi i \over M} l (m-n)} D^j_{m,n} \left( g \right) \,,}
and therefore
\eqn\calcptwo{
P^-_M(g) := {1\over M} \sum_{l=0}^{M-1} 
D^j_{m,n} \left(\Gamma_M^l \, g \, \Gamma_M^{-l}\right) = \left\{
\eqalign{ D^j_{m,n} (g) \qquad & \hbox{if $m-n \equiv 0\; ({\rm mod}\, M$)} 
\cr 0 \qquad & \hbox{otherwise.}} \right.}
Similarly, we have 
\eqn\calcpthree{
P^+_N(g) := {1\over N} \sum_{k=0}^{N-1} 
D^j_{m,n} \left(\Gamma_N^k \, g \, \Gamma_N^{k}\right) = \left\{
\eqalign{ D^j_{m,n} (g) \qquad & \hbox{if $m+n\equiv 0\; ({\rm mod}\, N$)}
\cr 0 \qquad & \hbox{otherwise.}} \right.}
Hence, if we replace $D^j_{m,n}(g)$ by 
\eqn\calcpfour{
P^+_N\,P^-_M(g) ={1\over MN} \sum_{l=0}^{M-1}\sum_{k=0}^{N-1} 
D^j_{m,n} \left(\Gamma_N^k \Gamma_M^l \, g \, 
\Gamma_M^{-l} \Gamma_N^k\right)}
we can actually drop the restriction \constraints\ on $m$ and $n$, and 
we arrive at the following expression for our symmetry-breaking boundary 
states
\eqn\gbstboson{
\gbs_{M,N} = 2^{-{1\over4}}\,\sqrt{MN}\;\sum_{j,m,n} 
    D^j_{m,n}\bigl(\,P^+_N\,P^-_M(g)\,\bigl)\;\ishjmn\,. }
We have made the radius $R={M\over N}R_{\rm s.d.}$ explicit for 
the moment, and have inserted the normalisation factor $\sqrt{MN}$ 
mentioned above. Upon computing the overlap between two such 
boundary states, each term in the $(k,l)$-sum from \calcpfour\ will 
give rise to an open string spectrum of the form \calcfive, and 
we obtain 
\eqn\genres{
Z_{g_1g_2}(\tilde q) = \sum_{l=0}^{M-1}\sum_{k=0}^{N-1} \sum_{n\in\Zop} 
\vartheta_{-{\alpha_{kl}(g_1,g_2)\over \sqrt{2}\pi} + \sqrt{2} n}
(\tilde{q}) }
where 
\eqn\alphadef{
2 \cos(\alpha_{kl}(g_1,g_2)) = \hbox{Tr}_{{1\over2}}\;
\bigl[\,g_1^{-1} \Gamma_N^k \Gamma_M^l \, g_2 \, 
                \Gamma_M^{-l} \Gamma_N^k\, \bigr]\,,}
the trace being taken in the fundamental representation $j=\half$. 
Equation \alphadef\ only determines $\alpha_{k,l}$ up to a multiple of
$2\pi$. However, the partition function $Z_{g_1g_2}$ is independent of
this ambiguity because of the summation over $n$. The 
partition function \genres\ defines a positive integer linear
combination of Virasoro characters, and thus Cardy's conditions are
met. 

The spectrum seems to exhibit `traces' of the underlying U(1) bulk
symmetry in a similar way as the spectrum of open strings stretched
between two non-coinciding  Dirichlet branes. But in contrast to the
latter situation, complete U(1) symmetry is not even restored in the
excitation spectrum of a single brane $\gbs_{M,N}$, for generic $M,N$:
if we consider the case for which $g_1=g_2=g$ with $g$ as in \group,
then \alphadef\ simplifies to  
\eqn\alphadefone{
\cos(\alpha_{kl}(g)) = 
 \cos\left({2\pi k \over N}\right)\, a^*a +
 \cos\left({2\pi l \over M}\right)\,b^*b\ .}
Clearly, the spectrum encoded in $Z_{gg}(\tilde q)$ is not, 
in general, invariant under addition of `U(1)-charges'. This shows 
once more that the boundary states $\gbs_{M,N}$ do indeed break the
U(1) symmetry.

In order to make contact with results obtained previously in the 
literature, let us specialise to SU(2) elements of the form 
\eqn\groupele{
g = g_\lambda := \pmatrix{ \cos\lambda & \sin\lambda \cr
- \sin\lambda & \cos\lambda}\,,}
and to the case $N=2,M=1$. Then the angles from above read 
\eqn\specialisti{
\alpha_{00}(g)=0\,, \qquad \alpha_{10}(g) = 2 \lambda +\pi\ ,}
and inserting these back into \genres\ reproduces the spectrum found 
in \refs{\rsone} in connection with truly marginal, symmetry-breaking 
deformations of a Dirichlet boundary state. 

Let us study the case $g_1=g_2=g$ further in order to establish an
interpretation of our family of boundary states $\gbs_{M,N}$.  First,
one notices that the spectrum $Z_{gg}(\tilde q)$ contains a single
vacuum state unless the elements of the SU(2) matrix $g$ satisfy
$b=0$ or $a=0$. Along those two curves in the 3-parameter space, the
vacuum occurs with $M$-fold (or $N$-fold) degeneracy.  This effect is
easy to understand since, for these values of $g$, our boundary states
can be expressed as superpositions of ordinary U(1) Dirichlet (or
Neumann) boundary states, $|\!|\, D,x_0\,\rangle\!\rangle$  (or 
$|\!|\, N,\tilde{x}_0\,\rangle\!\rangle$), at special (equidistant)
values of locations $x_0$ (or Wilson lines $\tilde{x}_0$); in our
conventions,
\eqn\superpos{\eqalign{
|\!|\,\smallmatrixone  
\, \rangle\!\rangle_{M,N} &= 
 \sum_{p=0}^{M-1} \; 
\bigl|\!\bigl|\,D,\; {\textstyle {2\pi R\;p\over M}}\,
          \bigr\rangle\!\bigr\rangle\ ,
\cr
|\!|\,\smallmatrix
\,\rangle\!\rangle_{M,N} 
&= \sum_{s=0}^{N-1}\;
\bigl|\!\bigl|\,N,\;{\textstyle {\pi\;s \over R\;N}}\,
              \bigr\rangle\!\bigr\rangle\ .  
\cr}}
As an aside, it follows by essentially the same observation that the
overlap between the boundary states labelled by $g\in {\rm SU(2)}$ and
the conventional U(1) Dirichlet and Neumann boundary states also give
rise to an integer linear combination of Virasoro characters in the
open string (and therefore that they also satisfy the mutual Cardy
conditions). 

On the other hand, there is no simple relation between the U(1)
boundary states and our boundary states for generic $g$ (\ie\ for
$ab\ne 0$). In this case, $Z_{gg}(\tilde q)$ always contains three
states of conformal dimension $h=1$: one from the U(1) vacuum
character $\vartheta_0(\tilde q)$ and one each from the two characters 
$\vartheta_{\pm\sqrt{2}}(\tilde q)$. For the cases $R=M R_{\rm s.d.}$
and $R = {1\over N} R_{\rm s.d.}$, the corresponding marginal boundary 
fields were studied in \refs{\rsone} and argued to be truly marginal
operators. In fact, one of these operators is associated to motions 
of the brane (or changes of the Wilson line), while the other two 
correspond to turning on boundary potentials as in \refs{\cklm,\polthor}. 
The situation at hand is technically slightly more complicated since 
one has to deal with boundary condition changing operators in order 
to study deformations of superpositions of Dirichlet or Neumann branes. 
Nevertheless, the main arguments used in \refs{\rsone} for the existence 
of a 3-parameter family of truly marginal operators should carry over 
to the general case $N\neq1,\; M\neq1$. We therefore conjecture that 
the marginal operators present here enjoy analogous properties, and 
that our family of boundary states is indeed connected by truly 
marginal deformations. Since, for generic $g \in {\rm SU(2)}$,  
the boundary conditions $\gbs$ are `elementary' in the sense that  
the vacuum occurs precisely once in $Z_{gg}(\tilde q)$, one is led 
to the conclusion that these deformations {\it merge} a set of $M$ 
equidistant Dirichlet branes into a single brane (of neither 
Dirichlet nor Neumann type, and preserving only the Virasoro algebra), 
until upon further deformation one reaches a system of $N$ Neumann 
branes with evenly spaced Wilson lines. 

One should, however, note that our parametrisation of boundary states 
by SU(2) elements chosen above is not quite one-to-one if $M\neq 1$ or 
$N\neq1$. From the definition of the representation matrices 
\matrixelem\ it follows that 
\eqn\ident{
D^j_{m,n}(g) = e^{-i\theta(m+n)}\; D^j_{m,n}(g(a_\theta)) = 
e^{-i\varphi(m-n)}\; D^j_{m,n}(g(b_\varphi))\,,}
where $g(a_\theta)$ is the matrix that is obtained from $g$ by
replacing $a$ by $e^{i\theta} a$ in \group, while $g(b_\varphi)$ is
the matrix that is obtained by replacing $b$ by $e^{i\varphi} b$. For
the boundary states we have constructed, only those matrix elements
arise for which $m+n$ is a multiple of $N$, and $m-n$ is a multiple of
$M$. Thus the brane that corresponds to $g$ is the same as the ones 
described by $g(a_\theta=2\pi{l\over N})$ or $g(b_\varphi=2\pi{l'\over M})$
for any integers $l,l'$. For the cases $M=1$ or $N=1$, the discrete 
identifications described above were overlooked in \rsone.
These are the only non-trivial identifications since \matrixelem\ 
also implies that 
\eqn\nontrivial{
D^j_{j,j}(g) = a^{2j} \ , \qquad D^j_{j,-j}(g) = b^{2j}\ ,}
so all other changes of $a$ and $b$ would be detected by the 
matrix elements that occur in the boundary states. 
Thus we conclude that the moduli space of our new boundary states at 
$R = {M\over N}~ R_{\rm s.d.}$ is given by the quotient  
\eqn\moduli{
{\cal M}_{M,N} = {\rm SU(2)} / \Zop_{M} \times \Zop_{N} \,.}

\newsec{The zero radius limit}

\noi
We can use the trick described in the previous subsection to
(re)derive the formula for the open string spectrum (between a brane and 
itself, say) in the zero radius limit (or similarly for $R \to \infty$). 
The zero radius limit 
corresponds to talking $N\rightarrow\infty$ while $M=1$. The relevant
projector that replaces \calcpfour\ is then given by the integral 
\eqn\infone{
{1\over \pi} 
\int_{0}^\pi d\theta\, D^j_{m,n} \left( \pmatrix{e^{i\theta} & 0 \cr 0 &
e^{-i\theta}} g  \pmatrix{e^{i\theta} & 0 \cr 0 & e^{-i\theta}}\right)
\,.}
(In order to deal with  $R \to \infty$, one simply replaces $\theta$ 
by $-\theta$ in the last matrix.) 
Up to an overall normalisation (which is formally infinite) the
relevant tree amplitude is then  
\eqn\calcin{
\A = {1\over \pi} \int_{0}^\pi d\theta\, 
\sum_{j\in\half\Zop_+} \Tr_{j} \left[
\pmatrix{e^{i\theta} & 0 \cr 
0 &e^{-i\theta}}\, g\,
\pmatrix{e^{i\theta} & 0 \cr 
0 & e^{-i\theta}} \, g^{-1} \right]\,\chi_{j^2}(q)   \;.}
For fixed $g$ as in \group, and for each value of $\theta$, the trace 
in \calcin\ can be written as  ${\sin((2j+1)\alpha)\over\sin\alpha}$ 
for some $\alpha\equiv\alpha(g,\theta)$. In order to determine $\alpha$, 
we work in the fundamental representation ($j=\half$), for 
which we obtain
\eqn\calcinone{\eqalign{
2 \cos\alpha & =\ {\rm tr} \left[
\pmatrix{  e^{2i\theta} a &  b\cr
- b^* &  e^{-2i\theta} a^* }\,
\pmatrix{ a^* & -b \cr b^* & a} \right] \cr
\noalign{\vskip5pt}
& =\; 2 \left[ 1 + a^*a\,(\cos(2 \theta) - 1) \right]\,.}}
Using trigonometric identities 
and taking square roots on both sides, \calcinone\ becomes
\eqn\calcinthree{
\sin\left({\alpha\over 2}\right) = 
|a| \sin \theta\,,}
and thus the angles in the partition function can be written as 
$\alpha=2 \pi f(g,\theta)$ with
\eqn\calcinfour{
f(g,\theta) = {1\over \pi}
\hbox{arcsin} \left( |a| \sin\theta \right)\,.}
Inserting this into our previous calculation, the open string 
spectrum is then given
by 
\eqn\inresul{
Z_{g,g} = {1 \over \eta(\tilde{q})} \sum_{n\in\Zop} 
\int_0^\pi {d\theta\over\pi}\,\, \tilde{q}^{(n + f(g,\theta))^2}
\,.}
In the special case $g= g_\lambda$ from \groupele, this agrees with the 
result obtained in \refs{\polthor} using fermionisation techniques.

\newsec{The factorisation property, completeness, and remarks on 
non-rational radii}

\noi
All the boundary states we have considered so far have the property
that, up to some irrelevant overall constant, the coefficients in front 
of the various Ishibashi states are given in terms of the matrix
elements $D^j_{m,n}(g)$ and therefore satisfy  
\eqn\factor{
D^{j_1}_{m_1,n_1}(g) \, D^{j_2}_{m_2,n_2} (g) 
= \sum_{j=|j_1-j_2|}^{j_1+j_1} C^{j}_{j_1,j_1}(m_1,n_1;m_2,n_2)\,
D^{j}_{m_1+m_2,n_1+n_2} (g) \,,}
where $C^j_{j_1,j_1}(m_1,n_1;m_2,n_2)$ are products of Clebsch-Gordan
coefficients (see \refs{\grwone} for more details). Indeed, the left
hand side of \factor\ is the matrix element between the states 
labelled by $(m_1\otimes m_2)$ and $(n_1\otimes n_2)$ in the tensor
product of the representations $j_1$ and $j_2$; the Clebsch-Gordan
coefficients describe the decomposition of this tensor product into
irreducible representations, which yields \factor. 

As follows from the detailed explanation in \refs{\grwone}, the
boundary structure constants 
$B^g_{(j;m,n)} \equiv {}^gB_{(j;m,n)}^{\bbbone}$ from the 
bulk-boundary OPE in the presence of the boundary condition $\gbs$ 
are given by $D^j_{m,n}(\iota \cdot g \cdot \iota^{-1})$, where
$\iota=\smallmatrix$.  
Therefore, eq.\ \factor\ has the same form as the sewing constraint
that arises from equating two different ways of evaluating a two-point
function of bulk fields on the upper half-plane: in one 
factorisation, the two bulk fields become arbitrary close to each
other and the leading term is determined by the bulk operator
product expansion, while in the other the two bulk fields come
separately arbitrarily close to the boundary and the leading term is
obtained from bulk-boundary operator product expansions. The
general form of this constraint, which can also be understood as a
cluster condition, is  
\eqn\cluster{
B^\alpha_a \,B^\alpha_b  = \sum_c \  \Xi_{abc}\,  B^\alpha_c
}
where (up to an $a$-independent factor) $B^\alpha_a$ is the
coefficient of the Ishibashi state associated  to the representation
$a$ in the boundary state labelled by $\alpha$  (in our case, 
$a=(j;,m,n)$), while $\Xi_{abc}$ is a product of the  
bulk OPE coefficient $C_{abc}$ with an element of the fusing matrix. 
In writing \cluster, the normalisation $B^\alpha_0 =1$ for the vacuum 
sector has been chosen; see  {\it e.g.}\ 
\refs{\Lew,\PSS,\FS,\BPPZ,\rsone} for more details. 
 
Comparison of this constraint with \factor\ suggests that in our 
case the $\Xi$ -- sometimes also referred to as the structure
constants of the `classifying algebra' \refs{\FS} --  are precisely
given by products of Clebsch-Gordan coefficients. As is explained in
detail in \refs{\grwone}, at the self-dual radius the structure
constants are indeed given by these group theoretic expressions. We
are only interested in the Virasoro symmetry here, and the different
representations are therefore completely characterised by their
conformal weight. Since these are independent of the radius $R$, we
can conclude that these structure constants do not depend on $R$, and
therefore that the boundary states we have constructed satisfy all
factorisation constraints that involve degenerate Virasoro
representations only. It seems plausible to us that the boundary
states also satisfy the other factorisation constraints (those
involving two non-degenerate representations and one degenerate
representation) provided that the group element $g$ is 
generic,\footnote{$^\ddagger$}{For non-generic group elements, \ie\ if  
$a=0$ or $b=0$, the self-overlap contains several sectors with
conformal dimension zero. Thus one does not expect the cluster
condition to hold in the same form, and it is obvious from \superpos\
that this will in fact not be the case. Compare also \refs{\rsone}.}
but we have not been able to check this directly.  
\sn
At the self-dual radius, it was shown in \refs{\grwone} that the most 
general solution to the factorisation constraint is parametrised by 
group elements in SL$(2,\Cop)$ rather than SU(2). In order to obtain 
the associated boundary states, one merely has to extend the
representation  
matrix $D^j_{m,n}(g)$ in \boundsu\ accordingly. The computations from 
\refs{\grwone} can be carried over to show that these additional 
boundary conditions again satisfy Cardy's condition; however, the 
open string spectra in general involve complex conformal dimensions. 
Furthermore, in a free field construction, some of these branes would 
appear to be localised at imaginary positions. For these reasons, one 
is inclined to discard those additional boundary states as `unphysical', 
at least for the purposes of string theory. 

Apart from these unphysical branes the boundary states we have
constructed above are the most general fundamental D-branes involving
only degenerate Virasoro representations. The only other Virasoro
Ishibashi states are in fact U(1) Ishibashi states. It therefore seems
very plausible that the above branes, together with the usual
Dirichlet and Neumann branes, from the complete set of 
{\it fundamental} conformal D-branes. 
 
It is generally thought that all other boundary states are
superpositions of such fundamental branes. In our case, however, 
we have to be more careful with the notion of superpositions than  
if we were to deal with a rational theory whose fundamental branes 
are given, say, by boundary states of the type constructed by Cardy 
in \refs{\cardy}. In that case, all further branes that are compatible 
with the Cardy states (\ie\ give rise to positive integer linear 
combinations of characters in all the open string spectra) lie in 
(a positive cone of) the lattice spanned by these fundamental 
D-branes. This can be shown with the help of the simple (and purely 
algebraic) completeness criterion first formulated in \refs{\PSS}. 
In the present case, the corresponding statements are somewhat more 
subtle since we have to deal with a continuum of fundamental boundary
conditions, and thus some analytic considerations come into play.    

As mentioned before, our fundamental boundary states consist of the
branes labelled by $g\in {\rm SU}(2)/\Zop_{M}\otimes \Zop_{N}$ with
$ab\ne 0$, together with the conventional Dirichlet and Neumann U(1)
boundary states (at arbitrary positions and Wilson lines). It is easy
to check that this collection of boundary states is `orthonormal' in
the sense that the vacuum is only contained  in the self-overlap of
such a boundary state (where it occurs precisely once, meaning that 
they are fundamental), but not in the spectrum of open strings
stretched between two different branes. Now let us consider an
arbitrary  conformal boundary state  
$|\!| B\,\rangle\!\rangle$ for the $c=1$ model at fractional radius
$R$. Every such boundary state can be expanded in terms of the
conformal Ishibashi states, 
\eqn\expandout{
|\!| B\,\rangle\!\rangle = \sum_{p} B_p |\, p,p \,\rangle\!\rangle
+ \sum_{w} B_w |\, w,-w \,\rangle\!\rangle + 
\sum_{j,m,n,} B_{(j;m,n)}\, \ishjmn  \,,}
where $B_p,B_w,B_{(j;m,n)}$ are some finite constants and 
$|\, p,p\,\rangle\!\rangle$ and $|\, w,-w \,\rangle\!\rangle$ are
the usual Dirichlet and Neumann Ishibashi states (with values of 
$p$ and $w$ appropriate for the radius $R$). We can then (at least 
formally) Fourier transform to write $|\!| B\,\rangle\!\rangle$
as
\eqn\fourier{\eqalign{
|\!| B\,\rangle\!\rangle = & 
\int_{S^1_{\!{1\over{2R}}}} d\tilde x_0 \; F_N(B;\tilde x_0)\; 
     |\!|\, N,\tilde{x}_0\,\rangle\!\rangle 
+ \int_{S^1_R} dx\; F_D(B;x_0)\; 
            |\!|\, D,{x}_0\,\rangle\!\rangle \cr
& \quad + \int_{\rm SU(2)'} d\mu(g)\; F_{\rm SU(2)}(B;g) \;\gbs\,,}}
where SU$(2)'={\rm SU(2)}/\Zop_M\times \Zop_N$ (excluding the cases
$ab=0$), and $d\mu(g)$ is the Haar measure. If
$|\!|B\,\rangle\!\rangle$ is again one of the branes from above, the 
$F_I$ are delta-functions. More generally, the expansion \fourier\ -- 
along with the calculations from Section~4 leading to the open string 
partition function -- make sense whenever the $F_I$ are {\it tempered 
distributions}. It therefore appears reasonable to take this as the
condition that characterises the appropriate generalisation of the
lattice of boundary states that arises for rational theories. In
particular, the (unphysical) solutions to the factorisation condition,
labelled by SL$(2,\Cop) {\hsmallsetminus} {\rm SU}(2)$, do not satisfy
this criterion.

\mn
Until now we have only considered rational radii 
$R= {M\over N} R_{\rm s.d.}$, including the limiting cases  
$R=0$ and $R=\infty$. For `non-rational' radii, the only Virasoro
Ishibashi states that correspond to degenerate representations arise
for $m=n=0$. The relevant matrix elements $D^j_{0,0}(g)$ then actually
only depend on $a$ and not on $b$. If we introduce $x \in [0,1]$ by 
$x=  2 |a|^2 - 1$, we find
\eqn\result{
D^j_{0,0}(g) = P_j(x) \,,}
where $P_j(x)$ is the $j$th Legendre polynomial. Indeed, it follows
from \matrixelem\ that the left hand side of \result\ is 
\eqn\specialist{\eqalign{
D^j_{0,0}(g) & = \sum_{l=0}^{j} {j\choose l} {j \choose j-l} 
\left(|a|^2\right)^{j-l} \left(-|b|^2\right)^l \cr
& = {1\over 2^j} \sum_{l=0}^{j} {j\choose l} {j \choose j-l} 
(x+1)^{j-l} (x-1)^{l} \,.}}
On the other hand, the Rodriguez formula for Legendre polynomials 
(see for example \refs{\arfken}) implies that 
\eqn\rodr{\eqalign{
P_j (x) & = {1\over 2^j j!}\left({d\over dx}\right)^j (x^2 - 1 )^j \cr
& = {1\over 2^j j!} \sum_{l=0}^j  {j\choose l} 
\left({d\over dx}\right)^l (x+1)^j 
\left({d\over dx}\right)^{j-l} (x-1)^j \cr
& = {1\over 2^j} \sum_{l=0}^{j} {j\choose l} (x+1)^{j-l} 
{j\choose j-l} (x-1)^{l}\,,}}
where we have used the Leibniz rule in the second line. The fact that
the coefficients of the boundary states for an irrational radius are
Legendre polynomials was already mentioned by Friedan \refs{\friedan};
an in-depth analysis of these boundary states (and in particular the
determination of the relevant overlaps) is in preparation
\refs{\romek}.\footnote{$^\star$}{We thank Romek Janik for sending us
copies of his notes prior to publication.}

\newsec{World-sheet supersymmetry}

\noi
We shall now exhibit boundary states similar to the $\gbs_{M,N}$ from 
above for the analogous supersymmetric situation. We start from a 
bulk theory that consists of a compactified free boson and a free 
Majorana fermion, and use only Ishibashi states from degenerate 
representations of the $N=1$ super Virasoro algebra. The gluing 
conditions we impose on the boundary states 
$|\!|B,\eta\rangle\!\rangle$ are 
\eqn\susygluing{
\bigl(\, L_n - \overline{L}_{-n}\,\bigr)\,
|\!|B,\eta\,\rangle\!\rangle
=\bigl(\, G_r + i \eta\;\overline{G}_{-r}\,\bigr)\,
|\!|B,\eta\,\rangle\!\rangle = 0\ ,
} 
where $G_r$ are the modes of the supercurrent, and $\eta$ is a sign to 
be discussed later. In addition to 
checking Cardy's condition, we shall have to address questions related 
to the GSO projection. Moreover, at least when passing to consistent 
ten-dimensional superstring backgrounds, it will be interesting to 
see whether the new boundary states are stable. 

The partition function of the $c={3\over2}$ bulk theory made up from a 
free boson and a free Majorana fermion is a product 
\eqn\susythree{
Z_{{\rm susy}} = Z_{{\rm boson}}\cdot 
  {1\over 2} \left\{ \left| f_3(q) \right|^2 + 
     \left| f_4(q) \right|^2 + \left| f_2(q) \right|^2 \right\} 
\,,}
of the partition function $Z_{{\rm boson}}$ from \part\ and that of 
the two-dimensional Ising model; the $f_i$ are functions defined by  
\eqn\fdef{\eqalign{
f_2(q) & = \sqrt{2}\, q^{{1\over 24}} \prod_{n=1}^{\infty} (1+q^{n})\ ,
\cr
f_3(q) & = q^{-{1\over 48}} \prod_{n=1}^{\infty} (1+q^{n-{1\over 2}})\ ,
\cr
f_4(q) & = q^{-{1\over 48}} \prod_{n=1}^{\infty} (1-q^{n-{1\over 2}})
\ .}}
The first two terms in \susythree\ describe the NS-NS sector (the sum 
of the two terms implements the GSO projection), while the last term
is the contribution from the R-R sector. For the above theory the 
GSO projection is $\half\bigl(1+(-1)^{F+\overline{F}}\bigr)$, where 
$F$ and $\overline{F}$ are the left- and right-moving fermion number 
operators, rather than 
${1\over 4}\bigl(1+(-1)^F\bigr)\bigl(1+(-1)^{\overline{F}}\bigr)$ as 
is familiar from the ten-dimensional superstring theories. The theory 
we are considering here is therefore the analogue of the type 0B (or 0A) 
theory, rather than of the type IIB (or IIA) theory \refs{\DixHar,\seiwitt}. 
It may be worth mentioning that, for $c={3\over2}$, it is inconsistent 
to GSO project using the other GSO projection, because then the fermionic 
(twisted) sectors one has to introduce in order to obtain a modular 
invariant partition function  do not obey level matching \refs{\DGH}.  

\subsec{The NS-NS sector}

Let us first concentrate on the NS-NS sector and collect some of the
relevant facts concerning the representation theory of the $N=1$
superconformal algebra ({\sl SVir}) at $c={3\over 2}$. In the NS
sector the degenerate representations have conformal weight
\refs{\cohnfried}    
\eqn\susyone{
h_{j} = {j^2 \over 2} }
with $j=0,1,2,\ldots$. Such a representation $\H_j^{{\rm SVir,NS}}$ 
has a null-vector at level $j+\half$ that generates the whole 
subrepresentation, and its character is therefore given by  
\eqn\susytwo{
\chi_{j}(q) = \left(\vartheta_{j}(q) - \vartheta_{j+1}(q)\right) 
f_3(q) \,.}
The only sectors of the free boson and fermion theory that give rise 
to degenerate NS-representations of the $N=1$ superconformal algebra 
are those with $p_L=({\hat{m}\over 2 R} + \hat{n} R ) = j \in \Zop$, 
and analogously for $p_R$. Thus, only the radii  
\eqn\susyfour{R= {M\over N} \,,}
where $M$ and $N$ are coprime integers give rise to degenerate 
{\sl SVir} representations. We also see that there is no radius where
all {\sl SVir} representations that occur in \susythree\ are
degenerate, the closest analogues of the self-dual radius from the
bosonic case being $R=1$ and $R={1\over 2}$, where half of the sectors
yield degenerate representations. 

For a radius as in eq.\ \susyfour, the momenta take the form 
\eqn\susyfive{
(p_L,p_R) = \left( {\hat{m}N\over 2M} + \hat{n}{M\over N}, 
{\hat{m}N\over 2M} - \hat{n}{M\over N} \right) 
=: (m,n)\,.}
The states with momentum and winding described by 
$(m,n)\in \Zop\times\Zop$ contain the representations
$\H_j^{{\rm SVir,NS}} \otimes\bar{\H}_j^{{\rm SVir,NS}}$ 
provided that $j$ is an integer with $j\geq |m|,|n|$. Thus the
possible Ishibashi states that involve these degenerate
representations are labelled by the triplets $(j;m,n)$ where $j$ is a
non-negative integer, and $m$ and $n$ are integers satisfying the
above inequality, and in addition by the sign $\eta= \pm1$ from 
\susygluing. These triplets correspond naturally to matrix elements 
of representations of $SO(3)$. As before we shall construct super 
Virasoro boundary states as linear combinations of the Ishibashi 
states $|j; m,n;\,\eta\,\rangle\!\rangle$, with coefficients given 
by $D^j_{m,n}(g)$. 

As in the bosonic case, there are $M$- and $N$-dependent conditions that
tell us which $(m,n)$ actually occur, {\it cf.} \constraints. Evaluating
the restriction $(p_L,p_R)\in \Zop\times\Zop$ carefully, one finds
that  only those degenerate {\sl SVir} Ishibashi states are contained
in the closed string theory for which the labels $m,\,n$ are integers
satisfying 
\eqn\superNSconstraints{\eqalign{
m+n &\equiv 0\;({\rm mod}\,N)\ \,,\quad\ 
m-n \equiv 0\;({\rm mod}\,2M)\quad\ \hbox{\rm if $N$ is odd}\,,
\cr
m+n &\equiv 0\;({\rm mod}\,N)\ \,,\quad\ 
m-n \equiv 0\;({\rm mod}\,M)\quad\quad \hbox{\rm if $N$ is even}\,.
\cr}}
Given that both $m$ and $n$ are integers, $m+n$ is even if and only if
$m-n$ is. Thus, distinguishing the cases is actually redundant in the
NS-NS sector, but will become necessary in the R-R sector. In order to
present all cases at once, we shall use the abbreviation 
\eqn\tildeMdef{
\widetilde{M} := \cases{2M \quad &if $N$ is odd$\,$, 
\cr  M \quad &if $N$ is even$\,$.\cr}}
The restrictions \superNSconstraints\ can be implemented with the help 
of the projectors \calcptwo\ and \calcpthree\ as before. Then the NS-NS 
parts of our $c={3\over2}$ boundary states can be written as 
\eqn\gbsNS{
\sgbs_{\rm NS-NS} = \sqrt{{\widetilde{M} N \over 2}}\,  \sum_{j,m,n} \ 
D^j_{m,n} \bigl(\,P^+_N P^-_{\tilM}(g)\,\bigr)\ 
|\,j;m,n;\,\eta\,\rangle\!\rangle \,,}
where $j,m,n$ are integers and $g\in {\rm SU(2)}$.

Before proceeding, we need to check whether the boundary state \gbsNS\ 
is invariant under the GSO projection
$\half\bigl(1+(-1)^{F+\overline{F}}\bigr)$. It is clear from \susythree\ 
that the momentum and winding ground states are GSO invariant. 
In order to determine the fermion number of the Virasoro Ishibashi
states, we consider the free boson and fermion representation with
highest weight state labelled by $(m,n)$, which coincides with the 
${\sl SVir}\times {\sl SVir}$ Verma module with highest weight 
$(h,\bar{h})=(\half m^2,\half n^2)$. The first singular vector with
respect to the left-moving {\sl SVir} occurs at conformal weight 
$(h,\bar{h})=(\half (m+1)^2,\half n^2)$, and it is the highest weight
state of an irreducible ${\sl SVir}\times {\sl SVir}$ representation
that occurs in the decomposition of the free boson and fermion
representation in terms of ${\sl SVir}\times {\sl SVir}$. Since the
left-moving conformal weights of the two vectors differ by a half-odd 
integer, they have opposite left-moving fermion number. Using this
argument repeatedly for left- and right-movers, we conclude that the
Ishibashi state corresponding to $(j;m,n)$ has eigenvalue
$(-1)^{2j-m-n}=(-1)^{m+n}$ under the GSO operator. Given
\superNSconstraints, it then follows that \gbsNS\ is indeed GSO
invariant.  

We have not specified the sign $\eta$ from our gluing conditions 
\susygluing\ yet. Indeed, the remarks in the previous paragraph show
that we can choose $\eta$ freely in the NS-NS sector: if 
$\gbs_{\rm NS-NS}$ is GSO invariant, then 
\eqn\othereta{
(-1)^F\, \sgbs_{\rm NS-NS}  = (-1)^{\overline F}\,\sgbs_{\rm NS-NS}}   
because the two operators commute and are of order two. It follows 
that $(-1)^F\sgbs_{\rm NS-NS}$ is another GSO invariant boundary state, 
which satisfies \susygluing\ with the opposite sign $-\eta$. As long 
as we focus on the NS-NS sector, we can therefore work with both sign  
choices. 

Using the same techniques as before, we can compute the overlap
between two such boundary states. For each $\alpha=\alpha_{k,l}$ that
shows up, {\it cf.}\ \alphadef, we have 
\eqn\susyeight{\eqalign{
\A^{{\rm NS-NS}} & = \half \sum_{j=0}^{\infty} 
{\sin((2j+1)\alpha) \over \sin\alpha} 
\left( \vartheta_{j}(q) - \vartheta_{j+1}(q)\right) f_3(q) \cr
& = \half \vartheta_0(q) f_3(q) + 
\half \sum_{j=1}^{\infty} \vartheta_j(q) f_3(q)
\left[ {\sin((2j+1)\alpha) \over \sin\alpha} 
- {\sin((2j-1)\alpha) \over \sin\alpha} \right] \cr
& = \half \sum_{j\in\Zop} \cos(2j\alpha)\, \vartheta_j(q) f_3(q) \,,}}
Upon a modular transformation, $f_3(-1/\tau)=f_3(\tau)$, and
therefore \susyeight\ becomes
\eqn\susynine{\eqalign{
\A^{{\rm NS-NS}} & = \half \int_{-\infty}^{\infty} dt\; 
\vartheta_t(\tilde{q})
f_3(\tilde{q})\ \sum_{j\in\Zop} e^{i(2\pi t + 2 \alpha)j} \cr
& = \half f_3(\tilde{q}) \int_{-\infty}^{\infty} d s\;
\vartheta_{s\over 2\pi}(\tilde{q})\
 \sum_{n\in\Zop} \delta(s+2\alpha +2\pi n) \cr
& = \half \sum_{n\in\Zop} \vartheta_{-{\alpha\over\pi} + n} (\tilde{q})
f_3(\tilde{q}) \,.}}
Up to the overall factor of $\half$ (which will be necessary in order
to implement the GSO projection in the open string sector), this
defines a trace over a positive integer number of representations of 
the $N=1$ NS superconformal algebra.

\subsec{The R-R sector}

Next we turn to the R-R sector. Here, the highest weight representations 
of the super Virasoro algebra can be characterised by the action of $G_0$ 
on the highest weight state, 
$G_0 |\lambda\rangle = \lambda |\lambda\rangle$. The corresponding
Verma module is degenerate provided that  
\eqn\susyrone{
\lambda = \pm \lambda_j := \pm {j \over  \sqrt{2}} \, }
for some $j\in\Nop_0+\half$. The conformal weight of the ground 
state is then given by 
\eqn\susyrtwo{
h^R_j = {j^2 \over 2} + {1 \over 16}\;.}
(Note that the `Ramond ground states' of the theory -- \ie\ the  
states with $h={c\over24}$ -- do not lie in degenerate 
representations of {\sl SVir}.) The representation with
$h_j$ as in \susyrtwo\ has a null-vector at level $j+\half$, and 
the corresponding character is therefore proportional to
\eqn\susyrthree{
\chi^R_j = f_2(q)\, \left(\vartheta_{j}(q) -
\vartheta_{j+1}(q)\right)\,.} 
Comparing \susyrtwo\ with the free boson spectrum \susyfive\ at 
radius $R= {M\over N}$,  one sees that a U(1)$\,\times\,$U(1) sector 
labelled by half-odd-integer $m,n \in \Zop + {1\over2}$ gives rise 
to a degenerate {\sl SVir}$\,\times\,${\sl SVir} representation 
$\H_j^{{\rm SVir,R}} \otimes\bar{\H}_j^{{\rm SVir,R}}$ for any 
$j \in \Zop + {1\over2}$ with $j\geq |m|,|n|$. The R-R part of the 
boundary states will therefore involve matrix elements $D^j_{m,n}$ 
of those SU(2) representations that do not define representations 
of SO(3). 

Again, we shall use projectors $P^+_N$ and $P^-_{\tilM}$ to
take care of the `gaps' that arise from the restrictions ({\it cf.}\
\superNSconstraints) 
\eqn\superRconstraints{
m+n \equiv 0\;({\rm mod}\,N)\ \,,\quad\ 
m-n \equiv 0\;({\rm mod}\,\widetilde{M})\,,
}
where now $m,n\in\Zop + {1\over2}$. The
R-R-part of our SU(2) family of {\sl SVir}-preserving boundary states
is thus given by 
\eqn\gbsR{
\sgbs_{\rm R-R} = \sqrt{{\widetilde{M} N\over 2}}\,  \sum_{j,m,n} \ 
D^j_{m,n} \bigl(\,P^+_N P^-_{\tilM}(g)\,\bigr)\ 
|\,j;m,n;\,\eta\,\rangle\!\rangle \,.}
This looks formally like \gbsNS, but now $j,m,n$ take half-odd-integer
values only.  

Before proceeding we need to analyse the action of the fermionic zero
modes (and the invariance under the GSO projection) carefully. Each 
momentum and winding number `ground state' occurs with two-fold
degeneracy since we have two fermionic zero modes $\psi_0$ and
$\overline{\psi}_0$ that anti-commute. Let us introduce creation and
annihilation operators  
$\psi^\pm ={1\over\sqrt{2}}(\psi_0\pm i\overline{\psi}_0)$
which satisfy the standard anti-commutation relations 
\eqn\psipm{ 
\{ \psi^\pm , \psi^\pm \} = 0\,, \qquad 
\{ \psi^\pm , \psi^\mp \} = 1 \,.}
We can then choose a basis for the highest weight space to consist of 
states $|(p_L,p_R),\pm\rangle$  that are (up to normalisation) uniquely
characterised by the conditions  
\eqn\ground{
\psi^{\mp} |(p_L,p_R),\pm\rangle = 0 \,,}
and we take the GSO operator to have eigenvalue $\pm 1$ on
$|(p_L,p_R),\pm\rangle$, irrespective of the value of
$(p_L,p_R)$. (This is the standard definition, compare for example
\refs{\bg}.) 

We are interested in boundary conditions that preserve the {\sl SVir} 
symmetry algebra only, and thus we need to rewrite the relevant
boundary condition in terms of $G_0$ and $\overline{G}_0$ rather than 
$\psi_0$ and $\overline{\psi}_0$. Let us therefore introduce the
operators $G^\pm = G_0\pm i\overline{G}_0$, which satisfy the 
anti-commutation relations 
\eqn\Gpm{
\{ G^\pm , G^\pm \} = 0\,, \qquad \{ G^\pm , G^\mp \} = 
2 (L_0 + \overline{L}_0) - {1\over 4} \,.}
For the momenta and winding numbers we are considering above, the
right hand side of the second equation is always non-zero (since $j=0$
is not allowed). Thus, for this set of representations, the space of
ground states can also be thought of as being generated by $G^\pm$. 
Depending on the choice of sign in the gluing conditions \susygluing, 
the Ishibashi states we are interested in are annihilated by either 
$G^+$ or $G^-$; in fact they satisfy 
$G^\pm |j;m,n;\pm \rangle\!\rangle = 0$.  
In either case, the Ishibashi state is (up to normalisation) uniquely 
fixed by this condition since it fixes the ground state component.   

On the ground states we have $G_0=\psi_0 p_L$ and
$\overline{G}_0=\overline{\psi}_0 p_R$. Our conventions therefore
imply that the state with GSO eigenvalue $+1$ is annihilated by
$G^-$ if $p_L=p_R$, but is annihilated by $G^+$ if $p_L=-p_R$. In 
particular, this implies that the Ishibashi states  
$|j;j,j;-\rangle\!\rangle$ and
$|j;j,-j;+\rangle\!\rangle$ are invariant under the GSO projection --
and those with the opposite signs are {\it not}. Taking into account the
relative fermion number of the various descendant {\sl SVir} singular
vectors (as before in the NS-NS case), we find that the Ishibashi
state $|j;m,n;\pm\rangle\!\rangle$ has GSO eigenvalue 
$\mp (-1)^{m-n}$. Given \superRconstraints\ it then follows that for
$N$ odd, all the relevant Ishibashi states $|j;m,n;-\rangle\!\rangle$
are GSO invariant, while for $N$ even, the same holds for the Ishibashi
states $|j;m,n;+\rangle\!\rangle$. This demonstrates that our boundary
states are GSO invariant as long as the suitable condition for the
zero modes is chosen; the latter is dictated uniquely by the sign
$\eta$ in \susygluing, so in contrast to the NS-NS sector there is only 
one family of GSO invariant R-R boundary states. 

Overlaps between such R-R boundary states can be computed in an almost 
identical fashion as before: for each $\alpha=\alpha_{k,l}$ we get  
\eqn\susyrfour{\eqalign{
\A^{{\rm R-R}} & = \half \sum_{j\in\Zop+\half} \cos(2j\alpha) 
\vartheta_j(q) f_2(q) \cr 
& = \half \int_{-\infty}^{\infty} \vartheta_t(\tilde{q}) f_4(\tilde{q}) 
\sum_{j\in\Zop+\half} e^{i(2\pi t+ 2 \alpha)j}\,,}}
where we have normalised the Ishibashi states so that they give rise to
$f_2(q)$ (rather than, say, ${1\over\sqrt{2}} f_2(q)$). We can rewrite
the last sum as  
\eqn\susyrfive{\eqalign{
\sum_{j\in\Zop+\half} e^{i(2\pi t+ 2 \alpha)j} 
& = e^{{i\over2}(2\pi t+ 2 \alpha)}\ 2\pi\;\sum_{n\in\Zop} 
  \delta(2\pi t + 2 \alpha + 2 \pi n) \cr
&= 2\pi \sum_{n\in\Zop} (-1)^n \delta(2\pi t + 2 \alpha + 2 \pi n)\,.}} 
Putting \susyrfive\ back into \susyrfour\ we then get 
\eqn\susyrsix{
\A^{{\rm R-R}} = \half  \sum_{n\in\Zop} (-1)^n
\vartheta_{-{\alpha\over\pi} + n}(\tilde{q})f_4(\tilde{q}) \,.}
This corresponds to the trace over (one half of) the NS sector trace
with the insertion of $\pm (-1)^F$. 

Finally, the whole boundary state is a linear combination of the NS-NS 
and the R-R component,
\eqn\susyboun{
|\!| g \rangle\! \rangle:= 
|\!| g,\eta\,\rangle \!\rangle_{{\rm NS-NS}} + i \,
|\!| g,\eta\,\rangle\!\rangle_{{\rm R-R}}\,\,,}
where the value of the parameter $\eta$ in the gluing condition for
the two contributions is the same. In view of GSO invariance in the 
Ramond sector, the sign depends on the radius $R={M\over N}$ as 
$\eta = (-1)^N$. Taking the various results together, we obtain 
\eqn\susytotal{
Z_{g_1g_2}(\tilde q) = 
\langle\!\langle g_1 |\!|\, q^{\half(L_0+\bar{L}_0)-{c\over 12}} 
|\!| g_2 \rangle\!\rangle = \sum_{k,l}\, 
\sum_{n\in\Zop} \vartheta_{-{\alpha_{kl}\over\pi} + n} (\tilde{q})\,
\half\bigl( f_3(\tilde{q}) - (-1)^n f_4(\tilde{q}) \bigr)}
for the spectrum of open strings attached to the branes \susyboun. 
This corresponds to a certain set of  GSO projected
NS-representations of {\sl SVir}. As is familiar from the analysis of
D-branes in type 0, the open string only consists of bosonic degrees
of freedom \refs{\bg}. Not all different open string sectors in
\susytotal\ have the same GSO projection, since the sign of the
GSO projection depends on $n$ mod(2). 
\sn
In addition to the boundary states \susyboun\ there exists another
class of boundary states that are given by 
\eqn\susybound{
|\!| \hat{g} \rangle\! \rangle:= \sqrt{2}\, 
|\!| g,\eta\rangle \!\rangle_{{\rm NS-NS}} }
where $|\!| g,\eta\rangle \!\rangle_{{\rm NS-NS}}$ is defined by
\gbsNS\ and where now $\eta= -(-1)^N$. Here we have used the fact, 
explained around eq.\ \othereta, that the NS-NS boundary states are 
GSO invariant for both choices of $\eta$. The open string spectrum 
of \susybound\ is 
\eqn\susytotald{
Z_{\hat{g}_1 \hat{g}_2}(\tilde q) = 
\langle\!\langle \hat{g}_1 |\!| \, q^{\half(L_0+\bar{L}_0)-{c\over 12}} 
|\!| \hat{g}_2 \rangle\!\rangle = \sum_{k,l}\, 
\sum_{n\in\Zop} \vartheta_{-{\alpha_{kl}\over\pi} + n} (\tilde{q})
f_3(\tilde{q}) \,,}
and therefore defines unprojected NS-representations of {\sl SVir}. 
Thus the boundary states \susybound\ are the analogues of Sen's
non-BPS branes from type II \refs{\sen} (see also \refs{\thompson}). 
The relative overlap between the two kinds of branes is also
consistent since we find 
\eqn\susytotaldd{
Z_{\hat{g}_1 g_2}(\tilde q) = 
\langle\!\langle \hat{g}_1 |\!| \, q^{\half(L_0+\bar{L}_0)-{c\over 12}} 
|\!| g_2 \rangle\!\rangle = \sum_{k,l}\, 
\sum_{n\in\Zop} \vartheta_{-{\alpha_{kl}\over\pi} + n} (\tilde{q})
{1\over \sqrt{2}} f_2(\tilde{q}) \,.}
Here we have used that the NS-NS overlap between two boundary states
with different values of $\eta$ gives rise to the same formula as
\susyeight\ except that $f_3(q)$ is replaced by $f_4(q)$; using the
modular transformation of $f_4$, we then arrive at \susytotaldd.
Given the normalisation \fdef\ of the function $f_2$, eq.\ 
\susytotaldd\ then describes an integer number of open string states
(which define R-representations of {\sl SVir}).
\sn
Similar to the bosonic case, the parameter space of the superconformal 
branes \susyboun\ is SU$(2)/\Zop_{\tilM}\times\Zop_N$.  
For special values of $g\in {\rm SU(2)}$, the branes are
superpositions of conventional Dirichlet or Neumann D-branes. For
example, for odd $N$, the brane that corresponds to the identity
matrix in  ${\rm SU(2)}$ can be written in terms of 
$2M=\widetilde{M}$ conventional Dirichlet branes that are 
equidistantly arranged around the circle and for which branes and
anti-branes alternate.\footnote{$^\ddagger$}{For the case $M=1$, this 
configuration is similar to the non-BPS D-brane that was constructed 
in \refs{\gutperle}.}  On the other hand, the brane that corresponds 
to the matrix  $\smallmatrix$ is described by $N$ (unstable) Neumann 
branes (whose Wilson lines take equidistant values) that are the 
analogues of Sen's non-BPS branes from type II \refs{\sen} (see 
also \refs{\thompson}). Indeed, for this value of $g$, only 
Ishibashi states with $n=-m$ contribute to the boundary state, and 
if $N$ is odd, none of these satisfies \superRconstraints; thus 
the boundary state has only a NS-NS component. (These non-BPS
D-branes are however different from the non-BPS D-branes in
\susybound\ since the boundary states satisfy different gluing
conditions corresponding to different values of $\eta$.)

For even $N$ the situation is reversed in that now the brane
\susyboun\ that corresponds to the matrix $\smallmatrix$ is described
by $N$ conventional Neumann branes and anti-branes (whose Wilson lines 
take again equidistant values such that branes and anti-branes
alternate), while the brane that is described by the identity matrix 
is described by $M$ unstable non-BPS Dirichlet branes (which are 
equidistantly localised around the circle). For two simple cases the 
relevant D-brane configurations are sketched in Figure~1.

\ifig\branespic{The brane configurations for $R=2/3$ corresponding 
to special group elements: on the left, we have $g = {\rm id}$ 
describing 2 Dirichlet branes $D$ and two anti-Dirichlet branes 
$\overline{D}$ distributed evenly over the target circle; on the 
right, $g = \smallmatrix$, describing three non-BPS Neumann branes 
$\hat N$ with Wilson lines that are evenly distributed over 
the dual circle.}{\epsfxsize4.8in\hskip-.5cm\epsfbox{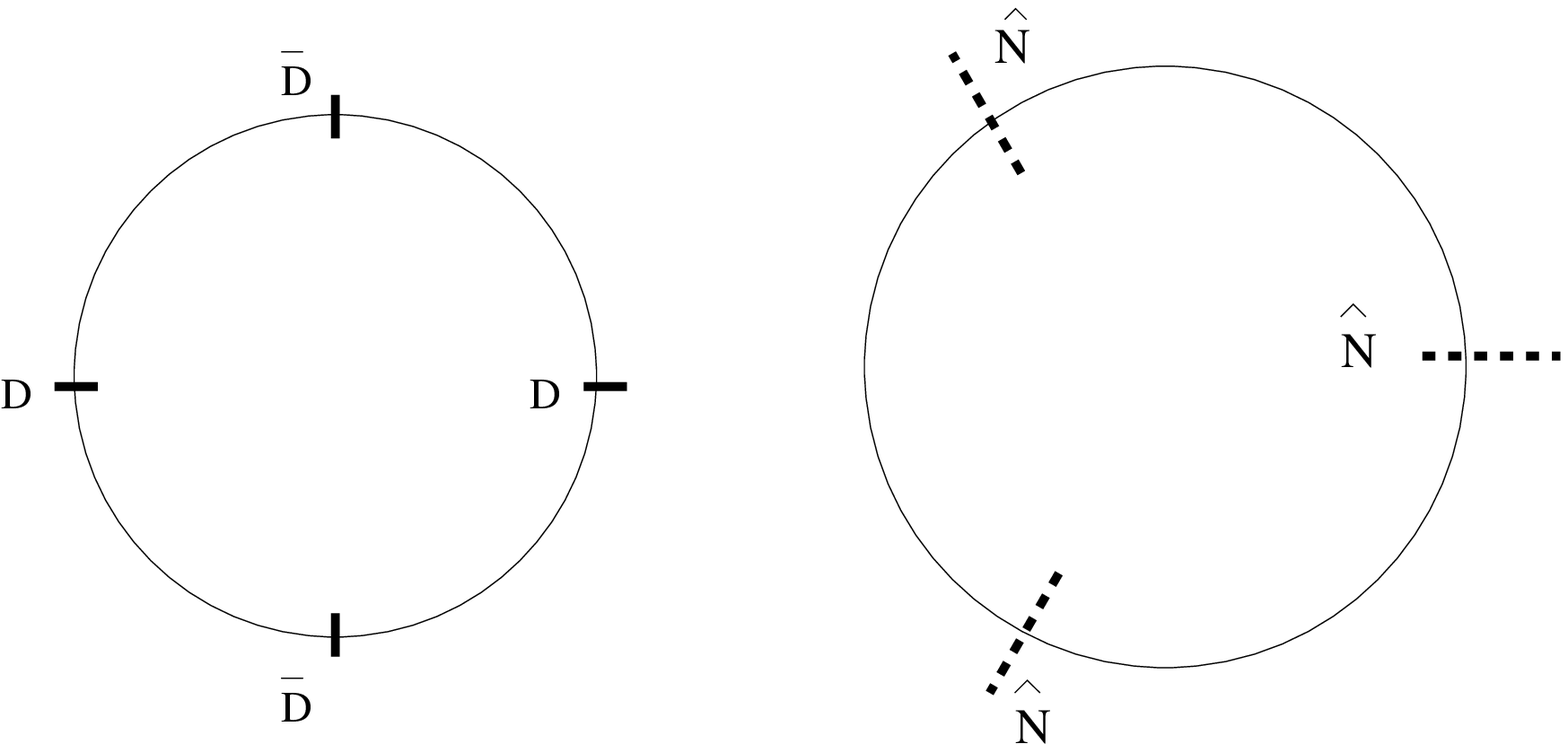}}

In either case, none of the branes carries non-trivial overall R-R
charge, since the zero-momentum R-R Ishibashi component (\ie\ the R-R
Ishibashi state with $m=n=0$) is absent. Thus it is consistent for
combinations of Dirichlet (or Neumann) branes to merge into non-BPS
branes. Similar statements also hold for the other branch of boundary
states that are described by \susybound; in this case, the space of
boundary states is ${\rm SO(3)}/\Zop_{\tilM}\times\Zop_{N}$.

\vskip.2cm
We have not attempted to check whether our superconformal boundary
states satisfy the factorisation constraint briefly discussed in
Section~6. The structural similarity with the bosonic case makes it
tempting to believe that the supersymmetric generalisations are  
again consistent, and form a complete set (apart from possibly
unphysical branes associated to 
SL$(2,\Cop)\hsmallsetminus {\rm SU}(2)$). However, the arguments from
\refs{\grwone} used in Section~6 rely on the fact that the branch of
the bulk moduli space that we were considering contains a rational
point where Cardy's boundary states are available, along with an SU(2)   
family of marginal deformations. These are all held to be fully 
consistent and can be used to determine the structure constants of
the classifying algebra, which do not depend on the circle radius. 
In contrast, the branch of the superconformal theories relevant 
for us does not have a point where the theory is rational with 
respect to an algebra containing {\sl SVir}. However, it may be
possible to exploit the fact that this branch is connected to the
super-affine line \refs{\DGH} where there is a point with $\su(2)_2$
symmetry. Similarly, one may be able to use this idea to make
statements about conformal boundary states for $c=1$ orbifold
theories. 

Apart from this technical difficulty, there is also the more
conceptual problem of what the relevant sewing relations for a
GSO projected superconformal field theory are. Indeed, the relevant
operator product in string theory must be GSO covariant, and therefore 
cannot simply be the naive conformal field theory operator product.

\newsec{Ten-dimensional boundary states, and stability}

\noi
We can use our boundary states from the $c={3\over2}$ theory to 
construct new boundary states for a full superstring background 
formed from ten copies of the free boson and free fermion system, 
roughly speaking by taking tensor products of the component 
boundary states. In contrast to ordinary Dirichlet branes, those 
boundary conditions break the U(1)$^{10}$ symmetry, but clearly this 
procedure will still produce a rather special class of all possible 
superconformal boundary conditions for a ten-dimensional toroidal 
target. We shall nevertheless write down the `tensor product' 
boundary states built from the states \susyboun\ from above 
since it is interesting to ask whether they are stable in the sense 
that no open string tachyons -- \ie\ boundary fields of total 
conformal dimension $h < {1\over2}$ --  occur in the spectrum. 
Moreover, the ten-dimensional setting  offers a second possibility 
for the GSO projection (leading to  type II string theory instead 
of type 0), and at the end of this section we shall discuss how 
to obtain GSO invariant boundary states for the type II theory. 

For the time being, we apply the same GSO projection 
${1\over2}\bigl(1 + (-1)^{F+\overline{F}}\bigl)$ as before, \ie\  
we work in a type 0 string theory (which is `plagued' by a closed 
string tachyon, but is otherwise consistent). Also, let us use the
light cone gauge and consider eight transverse  directions only. When
trying to build up eight-dimensional boundary states from those
discussed in the previous section by taking  
tensor products, one has to bear in mind that modular invariance 
of the eight-dimensional bulk theory forbids any mixing of 
NS with R sectors from different $c={3\over2}$ component theories. 
Therefore, the tensor product has to be taken in NS-NS and R-R sectors 
separately, and the boundary states we consider have the form 
\eqn\tensorprod{
|\!| \vec g\, \rangle\!\rangle := 
|\!| g_1 \rangle\!\rangle_{\rm NS-NS} \otimes \cdots \otimes 
|\!| g_8 \rangle\!\rangle_{\rm NS-NS} + i \;
|\!| g_1 \rangle\!\rangle_{\rm R-R} \otimes \cdots \otimes 
|\!| g_8 \rangle\!\rangle_{\rm R-R}\ .
}
The self-overlap $Z_{\vec g \vec g}(\tilde q)$ of such a 
boundary state can be calculated with the help of formulas 
\susynine\ and \susyrsix\ from the last section, and one 
finds
\eqn\tendimpf{
Z_{\vec g \vec g}(\tilde q) = {\cal A}^{\rm NS-NS}_\otimes
                           - {\cal A}^{\rm R-R}_\otimes
}
with 
\eqn\tensoverlap{\eqalign{
 {\cal A}^{\rm NS-NS}_\otimes &= {1\over2^8}\; f_3^8(\tilde q) \;
        \prod_{i=1}^8\; \sum_{n_i\in\Zop}\ \sum_{k_i,l_i}\;
            \vartheta_{-{\alpha_{k_il_i}\over\pi} + n_i} (\tilde{q})
\cr
 {\cal A}^{\rm R-R}_\otimes &= {1\over2^8}\; f_4^8(\tilde q)\; 
    \prod_{i=1}^8\; \sum_{m_i\in\Zop}\ \sum_{k_i,l_i}\; 
  (-1)^{m_i}\; \vartheta_{-{\alpha_{k_il_i}\over\pi} + m_i} (\tilde{q})\ .
\cr}}
All indices $i$ run from 1 to 8, and the range of $k_i$ and $l_i$ is 
determined by the radius $R_i$ as before. We have chosen the prefactor
of the Ramond part of the boundary state in \tensorprod\ so as to
eliminate the  vacuum state from the spectrum \tensoverlap, but it is
easy to see that, apart from the cases $a=0,\, M=1$ and $b=0,\, N=1$,
the open string spectrum still contains states whose total conformal
dimension is less than $\half$. For example, the
characters with $m_i=0$ for $i=1,\ldots,7$ and $m_8=\pm1$ occur with
the `wrong' GSO projection, and therefore \tendimpf\ contains a state
of conformal dimension ${1\over 2\pi^2} (\alpha_{k_8l_8} \pm \pi)^2$. This 
corresponds to an open string tachyon for a suitable choice of $k_8$ and 
$l_8$: for example, if $ab \neq 0$ we can choose $k_8=1$ and $l_8=0$
for $N_8>1$, or $l_8=1$ for $N_8=1$. In view of the fact that our
boundary states are related to superpositions of brane anti-brane
pairs, or to branes similar to Sen's non-BPS branes, it is not very
surprising that they are unstable.  
\mn
In the ten-dimensional case, we can also implement the more
restrictive GSO projection 
${1\over 4}\Bigl[ 1+(-1)^F)(1\pm(-1)^{\overline{F}}\Bigr]$; 
the resulting theory is then type IIB/IIA string theory. In 
the following we shall restrict ourselves to discussing the 
situation for IIB; the analysis for IIA is completely analogous.

It turns out that only relatively minor modifications in the spectrum 
of branes occur when passing from type 0 to type IIB, since the 
definition of boundary states has to be adjusted only slightly in 
order to insure invariance under the more restrictive GSO projection.
Suppose that some boundary state $|\!|B\,\rangle\!\rangle$, (for 
example $|\!| \vec g\,\rangle\!\rangle$ from before) is invariant 
under $(-1)^{F+\overline{F}}$. Since both $(-1)^F$ and  
$(-1)^{\overline{F}}$ are of order two, it follows that 
\eqn\obvious{
(-1)^F |\!| B\, \rangle\!\rangle 
   = (-1)^{\overline{F}}\, |\!| B\, \rangle\!\rangle\,.}
Thus the combination 
\eqn\susybound{
|\!| B\, \rangle\!\rangle^{\rm II} = {1\over\sqrt{2}} 
\Bigl(\,|\!| B\, \rangle\!\rangle + (-1)^F |\!| B\, \rangle\!\rangle
        \,\Bigr)}
is  invariant under the more restrictive GSO projection of the type 
IIB theory. This boundary state $|\!| B\, \rangle\!\rangle^{\rm II}$
is non-trivial if $|\!| B\,\rangle\!\rangle$ is, because 
$|\!| B\,\rangle\!\rangle$ cannot be proportional to 
$(-1)^F |\!| B\,\rangle\!\rangle$; this follows from the fact that 
application of $(-1)^F$ to the boundary state switches the sign $\eta$ 
in the  gluing condition \susygluing\ for the world-sheet fermions.  
On the other hand, different boundary states $|\!| B\,\rangle\!\rangle$  
from type 0 will lead to the same $|\!|B\rangle\!\rangle^{\rm II}$;
examples are the states $|\!| B\,\rangle\!\rangle$ and 
$(-1)^F |\!| B\, \rangle\!\rangle$ from the type 0 theory.  Thus we 
expect that in general there are more D-branes of type 0 than there are 
of type II, as is indeed the case for conventional D-branes \refs{\bg}. 

The bosonic part of the open string spectrum of 
$|\!| B\, \rangle\!\rangle^{\rm II}$ is precisely the same as that 
of $|\!| B\, \rangle\!\rangle$ and, therefore, the stability 
analysis is as before. But open strings between 
$|\!| B\, \rangle\!\rangle^{\rm II}$ branes will in addition have 
a Ramond sector. In order to see this, we first note that the overlap 
between $(-1)^F |\!| B_1 \rangle\!\rangle$ and 
$(-1)^F |\!| B_2 \rangle\!\rangle$ is the same as that
between $|\!| B_1 \rangle\!\rangle$ and $|\!| B_2 \rangle\!\rangle$. 
Furthermore, the overlap between $|\!| B_1 \rangle\!\rangle$
and $(-1)^F |\!| B_2 \rangle\!\rangle$ vanishes in the R-R sector 
(since the ground states satisfy different gluing conditions for the
zero modes), and in the NS-NS sector it leads to the expression from
above except that now $f_3(q)$ is replaced by $f_4(q)$. Under a
modular transformation, the functions $f_4(q)$ become $f_2(\tilde q)$,
and this gives rise to the R-sector of the open string.

\newsec{Conclusions and open problems}

\noi
In this paper we have constructed a family 
${\rm SU(2)}/ \Zop_{M}\times \Zop_{N}$  of conformal boundary
states for the theory of a free boson on a circle of rational
radius $R={M\over N}R_{\rm s.d.}$. These boundary states interpolate
between combinations of Dirichlet and Neumann branes, but are
fundamental, and symmetry breaking, for intermediate parameter
values. We have shown that the relative overlaps between arbitrary
elements of this family of boundary states satisfy Cardy's
condition. We have also demonstrated that each of them satisfies the
factorisation property at least for the subset of the bulk fields
that correspond to degenerate Virasoro representations. Finally, we
have proposed an analytic characterisation of the space of conformal
boundary conditions that are spanned by these branes together with 
the usual Dirichlet and Neumann branes. 

We have argued that the marginal operators that deform the branes
along the moduli space are in fact truly marginal; it would be
interesting to check this directly. It would also be interesting to
verify that the branes we have constructed satisfy the other sewing
constraints such as associativity of the boundary operator product
expansion or, for that matter, the factorisation property for the 
non-degenerate bulk representations. One would then expect to be 
able to determine the most general (fundamental) boundary state 
along the lines of \refs{\grwone}. In fact, given the results of 
\refs{\grwone}, it seems very plausible that the most general 
boundary conditions that satisfy the factorisation and Cardy 
constraints are described by the same formula we used here, but 
with $g$ being an arbitrary group element in SL$(2,\Cop)$ rather 
than in SU(2). The branes that are associated to 
SL$(2,\Cop)\hsmallsetminus {\rm SU}(2)$ do not satisfy the criterion 
discussed in Section~6, as the $F_I$ appearing in an expansion in 
terms of our fundamental branes do not define distributions on the 
space of $C^\infty$-testfunctions. However, they also have
imaginary couplings to the bulk fields, and are presumably not
actually relevant in string theory. In this sense the branes that we
have discussed in this paper are likely to form a complete spanning
set for the physically relevant boundary states. 

The boundary states that we have constructed depend critically on the
precise value of the radius. (This simply reflects the fact that the
same is true for the set of Virasoro Ishibashi states.) It would 
therefore be interesting to understand what happens to the boundary 
states as the radius is varied. One may guess that two of the marginal 
operators that are present for rational radii become relevant as the 
radius becomes irrational, and that the intermediate boundary states 
decay to combinations of conventional Dirichlet or Neumann boundary 
states, but the details remain to be worked out. The fact that these
branes exist only for a discrete set of radii suggests that they do
not have a straight-forward geometrical interpretation, but it would
nevertheless be interesting to explore this issue.

We have also considered the theory of a free boson and a free fermion
on a circle with $N=1$ world-sheet supersymmetry, and we have constructed 
a family of boundary states that preserve the corresponding super 
Virasoro algebra. While for the `one-dimensional' case our boundary 
states are likely to form a complete set, the D-branes we have 
constructed for the ten-dimensional theory satisfy very special 
gluing conditions and are far from describing the most general case. 
(In fact, the D-branes we have constructed preserve the eight separate 
super Virasoro algebras associated to each of the eight directions in 
light-cone gauge, while in general only the diagonal super Virasoro 
subalgebra needs to be preserved.) One should try to investigate the 
general situation further. 

Another extension worth studying is that to $N=2$ world-sheet 
supersymmetry. All of the branes we have constructed are non-BPS and 
in fact unstable. This is not surprising since we have only preserved 
$N=1$ supersymmetry on the world-sheet, while one presumably needs to 
have $N=2$ supersymmetry on the world-sheet in order to obtain spacetime
supersymmetric D-branes. It should therefore be worthwhile to analyse
whether our construction can be generalised to the case where one
preserves an $N=2$ super Virasoro algebra.  

\vskip 1cm

\centerline{{\bf Acknowledgements}}\pano

\noindent We are grateful to G\'erard Watts for collaboration at an
early stage of this work, and for many helpful comments and
discussions. We also thank Romek Janik, Juan Maldacena, Daniel
Roggenkamp, Katrin Wendland and Jean-Bernard Zuber for useful
conversations. M.R.G.\ is grateful to the Royal Society for a
University Research Fellowship, and the research of A.R.\ is supported
in part by the Nuffield Foundation, Grant Number NUF-NAL/00421/G. This
work is also partly supported by EU contract HPRN-CT-2000-00122. 

\vskip1cm
\footatend
\immediate\closeout\rfile\writestoppt
\baselineskip=14pt\centerline{{\bf References}}\bigskip{\frenchspacing%
\parindent=20pt\escapechar=` \input refs.tmp\vfill\eject}\nonfrenchspacing

\bye